**Main Manuscript for**

# Technology Creates a New Synchrony Advantage in Global Scientific Collaboration.

Zheng Fu[a,1,2] and Zhuofan Li[b,1,2]

[a]School of Sociology, University of Arizona

[b]Department of Sociology, Virginia Tech

[1]To whom correspondence may be addressed.

**Email:** zfu@arizona.edu; zhuofan@vt.edu.

[2]Z.F. and Z.L. contribute equally to this work.

**Author Contributions:** Z.F. and Z.L. designed research. Z.F. and Z.L. performed research. Z.F. and Z.L. analyzed data. Z.F. and Z.L. wrote the paper.

**Competing Interest Statement:** The authors declare no competing interest.

**Classification:** Social Sciences, Social Sciences

**Keywords:** scientific collaboration; technology; time; remote work; inequality

**This PDF file includes:**

Main Text
Figures 1 to 5
Tables 1 to 1

**Abstract**

Technological progress has made remote collaboration easier than ever, but its implications for inequality in global science remain unclear. While messaging apps, videoconferencing software, cloud computing, and online collaboration platforms now help scientists work together across geographic distances, they may also increase our expectations for synchronous interaction, creating a new *synchrony advantage* for scientists who share more working hours with their collaborators. The pandemic-induced expansion of videoconferencing provides an opportunity to examine the emergence of this synchrony advantage. Drawing on publication data from 846,618 researchers affiliated with 496 leading research institutions worldwide between 2015 and 2025, spanning five scientific domains and 18 time zones, we compare international collaborations between pairs of authors and institutions with greater and lesser working-hour overlap before and after 2020. We find a gap in both the extent and output of collaboration between higher- and lower-overlap pairs, which was stable before the pandemic but widened significantly after. Robustness tests rule out a series of alternative explanations and Covid-related shocks that could have produced the same gap. This synchrony advantage is stronger in fields that rely more heavily on real-time coordination and imposes the heaviest burden on scientists in Oceania, South and West Asia, and Latin America. These findings suggest that digital technologies do not merely fail to eliminate geographical barriers, but create new axes of inequality in global scientific collaboration, with implications for how to best support scientists and teams across time zones.

**Significance Statement**

Technological progress has made remote collaboration easier. This study shows that it has paradoxically created a new axis of inequality around time differences. After the pandemic made videoconferencing one of the default modes of scientific collaboration, researchers whose working hours overlap more with their collaborators gained a new persistent advantage in forming and sustaining international collaborations, while those separated by multiple time zones fell behind. As science becomes increasingly organized around synchronous interactions, our findings show the importance for the scientific community to understand the effect of technology as conditioned by the temporal organization of scientific tasks and to facilitate travel and build infrastructure that supports researchers working across time zones.

## Main Text

### Introduction

Since the rise of the Internet, scientists have hoped that technology would make international scientific collaboration easier and eventually democratize how we do science (1). Decades later, that hope seems fulfilled, at least in part. Email, messaging apps, videoconferencing, cloud computing, and online collaboration platforms allow scientists to work together across geographic distances. International collaboration became increasingly prominent, for example in the U.S. rising from 23% of science and engineering publications in 2003 to 40% in 2022 (2). However, the world of science is far from flat. Scientific productivity and scientific impact remain concentrated in relatively few developed countries and elite institutions, and stratified by proximity to this core (3–10). This persistence of inequality raises the question: does technology merely fail to eliminate existing barriers, or can it also create new ones?

Time is one such barrier. As technology makes distance matter less, it can paradoxically make time matter more. Research on remote collaboration has long recognized that time differences create coordination challenges distinct from ones caused by physical distance (11–15). While digital technologies facilitate remote communication asynchronously, they do not eliminate the temporal constraint on synchronous interactions — having a conversation in real time still requires collaborators to be available at the same time (14). Moreover, when a new technology becomes available, collaborators do not merely fit it into their existing workflow, but often change the norms and expectations around their collaboration. Being able to check emails on your mobile devices, for example, offers individuals greater flexibility in when and where to work, but has also fostered a collective norm of remaining constantly available (16). Now, if scheduling a Zoom call has become the routine, why wait for a reply by email?

We argue this growing expectation for synchronous interactions induced by digital technology can make time a new axis of widening inequality. Scientists separated by great geographical distance, for example, are now expected to connect more than ever before, but when it comes to scheduling a Zoom meeting, they still have far fewer working hours to choose from than colleagues who work from the same time zone (14). Even if scientists are willing to take the bad hours, a lack of overlap in working hours still creates a structural barrier to sustainable and productive collaboration in the long term. As scientists increasingly move seminars, conferences, meetings, writing sessions, and even lab meetings online, has the shift to synchronous collaboration widened the inequality in scientific collaboration between those who share more working hours and those who share fewer?

The expansion of videoconferencing during the COVID-19 pandemic offers an opportunity to test whether technology gives some scientists a “synchrony advantage.” On the one hand, the adoption of videoconferencing induced by global lockdowns was abrupt enough that we can treat it as an exogenous shock not driven by individual or institutional decisions (17, 18). Within weeks, travel stopped, laboratories closed, and videoconferencing largely became the default mode of collaboration. By April 2020, 37% of Americans worked from home, compared with 5% before the pandemic (11), while Zoom use increased 30 times (19). On the other hand, the shock did not alter pre-existing time differences between potential collaborators, leaving any overlaps in their working hours unchanged. These conditions allow us to exploit the interaction between the shift to videoconferencing and any pre-existing time differences in a difference-in-differences design: we compare the extent and output of international collaboration, before and after the lockdown, between pairs of collaborators with more and less overlap in their working hours.

We analyze international collaboration among 846,618 authors from 496 leading academic research institutions, drawing on 1,401,721 co-authored publications between 2015-2025 that span five fundamental domains of science and 18 time zones. We find that the gap in both the extent and output of collaboration between authors and institutions with higher versus lower working-hour overlap was stable before the pandemic, but started to widen after 2020. We consider a series of alternative explanations that could have produced the same gap, including other Covid-related shocks, and find this widening gap to most likely reflect a synchrony advantage as the result of videoconferencing. This synchrony advantage is unevenly distributed across fields and continents, with a disproportionate burden on fields organized around interdependent, real-time coordination, and Australia, New Zealand, most of Asia compared to Europe and Latin America.

This paper makes four contributions. First, it extends research on remote work, which largely focuses on where and how people collaborate, by showing when collaborators are able to work together also influences collaboration, since scientists is one of the professional groups with the most flexible schedules, this constraint might be even more pronounced in other professions. Second, it highlights how the effect of technology is mediated by the temporal arrangement of scientific work and the continuing significance of real-time coordination. Third, it implicates heterogeneity across fields, locations, and power asymmetry to shed light on potential mechanisms and its interaction with reliance on real-time coordination. Last, it identifies how digital technologies, instead of mitigating geographical inequality, adds a temporal axis to longstanding core-periphery inequalities in knowledge production.

## Research Design

Figure 1 illustrates our research design. We define “working-hour overlap” as the number of hours two collaborators share within a normal working day of 8 AM to 6 PM local standard time, following the convention in the literature (12–14). Although scientists may work beyond conventional working hours, it is still easier to schedule videoconferences when collaborators have more overlap.

Our sample consists of 1,401,721 co-authored publications between 496 leading academic research institutions from 2015 to 2025, spanning five fundamental domains of science: life sciences, physical sciences, psychological and cognitive science, social science, and economic and management science. We restrict the sample to cross-country collaborations and publications with 30 or fewer authors. We construct two panels. The author panel tracks 846,618 authors and their collaborators in each of 18 time zones. The institution panel tracks 80,653 cross-country institution pairs of which 63,509 collaborated at least once. For both panels, we also exclude pairs with more than 7 hours of overlap, as these are typically in the same or neighboring countries where other forms of proximity, such as geographical and cultural, dominate how scientists collaborate. We measure three outcomes at the author level — whether a new collaboration forms, how many new collaborators appear, and how many publications result — and the latter two at the institution level as most institutions already collaborate.

We treat the pandemic-induced shift to video conferencing as an exogenous shock to synchronous collaboration. Although the shock was common, its effect depends on how many working hours a pair shares. We expect a new technology that facilitates synchronous interactions to benefit high-overlap pairs more than low-overlap. In the absence of a shock, collaboration between high-overlap and low-overlap pairs would have evolved similarly on average, conditional on separable changes to both collaborators’ time zones and institutions (20). While a global pandemic is too messy to support clean causal identification on its own, our design traces the emergence of the synchrony advantage net of time-invariant heterogeneity and of time-varying shocks common to a collaborator time zone or, in the institution-pair design, to either institution. We test a series of alternative model specifications and causal mechanisms in SI Appendix (Table S3-S12 and Figs S3-S13), including a binary treatment (high versus low overlap) and a subsample of papers that do not span three or more time zones. Details on data construction, data validation, outcome variables, model specifications can be found in Methods and Materials.

## Collaboration Shifted Toward Pairs with More Shared Working Hours Post-Covid

Until the pandemic, collaborators with more overlap in their working hours had no systematic advantage over those with less. High-overlap and low-overlap pairs followed parallel trends across all four outcomes. If anything, it was the high-overlap pairs that fell slightly behind on most author-level outcomes. Figure 2 plots the raw trends in collaboration for high-overlap (5–7 hours) and low-overlap (0–2 hours) pairs from 2015 to 2025. Across all four outcomes, the gap between high- and low-overlap pairs was flat or slightly negative until 2020. Starting in the first full year after global lockdowns, the trends reverse. High-overlap pairs start to outpace low-overlap pairs, and the gap widens through 2025. This divergence coincides with the pandemic-induced shift to synchronous interactions through videoconferencing. In the post-treatment period (2021–2025), high-overlap pairs grew faster than low-overlap pairs on most outcomes. At the author level, the number of collaborators grew by 51% for high-overlap pairs versus 32% for low-overlap pairs; publications grew by 54% versus 34%. At the institution-pair level, collaborating author-pairs grew by 51% versus 31%. The gap is smaller for institution-pair publications, where both groups grew by about 40%, before controlling for institution-level confounds. Table S0 presents full descriptive statistics and lists county pairs that account for the largest share of collaborations at each hour. Low overlap (0-2 hours) is dominated by trans-pacific and Eurasian ties, while high overlap (5-7 hours) includes collaborations within regions, like Canada-US, Australia-China, and also north-south ties across long distances, such as Brazil-Germany, Brazil and UK (Table S13).

**Difference-in-Differences**. Consistent with the descriptive evidence, our difference-in-differences estimates show that after the shift to videoconferencing, the gap between high-overlap and low-overlap pairs widens significantly in both the extent and the output of international scientific collaboration. Table 1 shows that at the author level, relative to 2015–2019, each additional hour of working-hour overlap is associated with an additional 0.29-percentage-point advantage in the probability of forming a new collaboration in that time zone (Table 1, column 1), 3.7% in the number of collaborators (column 2), and 2.8% in collaborative publications (column 3) favoring higher-overlap pairs, all significant at the 0.005 level. At the institution-pair level, the gap per hour of overlap widens by 4.0% for collaborating researcher pairs (column 4) and by 2.6% for co-authored publications (column 5), both significant at the 0.005 level. These estimates are robust to alternative specifications which we explore in SI Appendix (Fig. S1 and Table S1).

**Event Study**. Did the widening already start before the pandemic? If the synchrony advantage can be attributed, at least in part, to the post-pandemic shift in expectation for synchronous interactions, we would expect the gap between higher- and lower-overlap collaborators to remain stable through 2020 and start to widen only thereafter. Figure 3 shows event studies consistent with a post-pandemic shift at both the author and institutional levels: parallel trends pre-COVID and widening gaps thereafter, net of the same fixed effects. Across all five outcomes, the pre-COVID coefficients (2015–2018) are small, stable, and statistically indistinguishable from zero, indicating that high- and low-overlap pairs were on comparable trends before the pandemic. The trends start to diverge in 2021: from there the gap per additional hour of overlap turns positive in every post-COVID year and keeps widening through 2025. This pattern — a flat pre-trend followed by a widening rather than one-time divergence — is consistent with our interpretation that the value of shared working hours increases only after the adoption of videoconferencing technology inflated expectations for synchronous interactions, and that the synchrony advantage has continued to grow in the years since.

**Placebo Tests.** As an additional test for the timing of the shock, we re-estimate our models on the pre-pandemic period (2010–2019), assigning a placebo shock to 2015, 2016, and 2017. All placebo estimates are small, statistically insignificant, and mostly negative across the five outcomes, in contrast to the positive and significant post-2020 estimates (Table S2, Panel B).

Another concern about the timing comes from publication delays: because journal articles appear one to two years after the underlying research, publications in 2021 could partly reflect collaboration that begun before the pandemic and bias our estimates downward. We re-estimate our models on the post-pandemic period by delaying the onset of the shock to 2022 or 2023, which yield estimates that are similar and even larger than post-2020 estimates (Table S2, Panel C).

### The Widening Gap Likely Reflects a Synchrony Advantage

While our main estimates uncover a robust widening gap between higher-overlap and lower-overlap collaborations that persisted throughout the post-COVID years, the expansion of Zoom use was not the only shock that the pandemic might have introduced. We argue that videoconferencing is the most likely causal mechanism because the conditions under which a competing mechanism could have generated the observed gap are stringent. It must (1) begin with the 2020 shock under the parallel trend assumption, (2) persist through 2025 after travel, schooling, and in-person work had resumed, and (3) affect a pair of collaborators or time zone at a distance simultaneously such that it is not absorbed by the pairwise and year fixed effects.

We explore several alternative mechanisms that are closest to meeting these conditions and could have produced a widening gap between high- and low-overlap pairs, without a technology-mediated synchrony advantage in having more working hour overlap.

**Covid-related Shocks.** While videoconferencing was not the only shock through which the pandemic disrupted scientific collaboration, it was the primary, if not the only, one to stay. Most of Covid-related shocks were transient and their effects should have subsided as travel, schools, and laboratories returned to normal by the end of 2022. Recall that our difference-in-differences estimates are largest in 2024-2025 (Figure 3) and in the placebo test with a delayed treatment, even larger than our baseline estimates (Table S2 and Fig. S2). This persistence is consistent with the expansion of videoconferencing as an abrupt yet durable change in science rather than a temporary disruption.

We additionally consider a series of more specific shocks. A shift toward smaller teams could mimic a synchrony advantage if COVID disproportionately broke up large teams spanning many time zones; in fact, average team size continued to rise after 2020 (from 5.1 to 5.4), and a triple difference on authors' pre-pandemic team size shows no differential widening, with a null Overlap × Post × Team size interaction for all outcomes (Table S3 and Fig. S3). Burden from additional caregiving due to school closure, which fell disproportionately on women and competed for the early and late hours on which low-overlap collaboration relies (21), should likewise have attenuated with school reopenings and affected female authors more than male authors; instead, the synchrony advantage persists through 2025 and is nearly identical for female and male authors, with null Overlap × Post × Female interactions for all three outcomes (Table S4 and Fig. S4). Nor is the widening an artifact of authors relocating closer to their collaborators: restricting the sample to 81% authors whose home time zone never changes over 2015–2025 leaves the estimates essentially unchanged (Table S5 and Fig. S5).

**Reallocation of Effort.** Because the same author or institution can appear in both high- and low-overlap pairs, our difference-in-differences estimates cannot, by design, rule out the possibility that authors and institutions simply shift effort from low- toward high-overlap collaborators without any additional return to working-hour overlap. While a conservative interpretation of the widening gap still points to a synchrony advantage, we would like to know whether it comes with a real benefit for high-overlap collaboration. Robustness tests show that the widening gap persists after restricting the sample to authors who have only ever collaborated with a single collaborator time zone and therefore cannot reallocate without finding new collaborators (Table S6 and Fig. S6), and after controlling for the number of alternative collaborators each author can choose, which neither moves the overlap estimates nor themselves predict any post-2020 decline in low-overlap collaboration (Table S7 and Fig. S7). These findings suggest high-overlap collaboration grew, rather than merely drawing time away from low-overlap collaboration.

**Geopolitical Tensions Between the US and China.** Among low-overlap collaborator pairs, a notably high proportion involves authors and institutions in mainland China: 46.0% of low-overlap author pairs and 62.5% of institutional pairs, compared to 18.8% and 23.0% for high-overlap collaborations). After 2020, collaboration with China could have shifted due to factors unrelated to the synchrony advantage, such as rising geopolitical tensions between the U.S. and China over COVID-19 that affected international collaboration. A robustness test shows that except for one institution-level outcome, the synchrony advantage remains robust after excluding all collaborations involving UTC+8 institutions (Table S8 and Fig. S8).

### The Synchrony Advantage Is Not Driven by Senior Scientists

Synchronous collaboration is not necessarily symmetric. When collaborators differ in status, the burden of synchronous interactions falls unevenly. On the one hand, junior authors may accommodate senior collaborators by meeting outside their own regular working hours and attenuate the effect of synchrony advantage for the most unequal collaborations. On the other hand, if senior authors — who enjoy greater autonomy over their own schedules and research agendas — respond to videoconferencing by choosing collaborators who could accommodate their working day, the gap would widen most where power is most unequal, and our pooled estimate would partly reflect status-driven selection rather than a synchrony advantage. Status therefore may shape not only whether remote collaborations form, but also who bears the temporal cost of sustaining them.

To investigate these possibilities, we estimate triple-difference models that interact Overlap × Post with within-pair status asymmetry, measured by the share of corresponding authorship an author or institution occupies on their pre-2020 co-authored publications. The triple interaction is small and statistically indistinguishable from zero across all outcomes and the DiD interaction Overlap × Post remains largely identical to the baseline at mean asymmetry (Table S9 and Fig. S9). In other words, the synchrony advantage widens as much between equal collaborators as between unequal ones. In addition to ruling out status-driven selection, this finding also helps validate our operationalization of overlap, which assumes that collaboration relies on standard working hours at both ends. If junior scholars routinely work outside their regular hours, effective synchrony would exceed measured overlap most where status asymmetry is greatest. We find no evidence of such accommodation.

### The Synchrony Advantage Is Concentrated in Fields That Require More Real-Time Coordination

In theory, science is a flexible job: many tasks, such as defining problems, analyzing data, and writing papers, do not require an immediate response. However, the relative flexibility compared to other occupations does not mean one's schedule is unconstrained in collaboration (22). Media synchronicity theory distinguishes conveyance, the transmission of information, which asynchronous media handle well, from convergence, the development of shared understanding, which benefits from synchronous interaction (23). Fields differ in how much of their work requires convergence. Where tasks require an interdependent workflow and collaborators to go back and forth during execution, we would expect working hour overlap to carry a strong advantage. In contrast, where tasks can be cleanly divided, sequenced, or pooled, collaborators can often wait to exchange output asynchronously (24). Laboratory sciences depend on tacit knowledge and on real-time troubleshooting of experiments (25), and on aligning the rhythms of instruments, organizations, and individual careers (26). Economics, by contrast, has a long-established working-paper culture organized around asynchronous exchange. We therefore expect the synchrony advantage to be largest in fields where collaboration depends most on convergence.

Is the synchrony advantage heterogenous across scientific fields with different temporal structures of tasks? Figure 4 decomposes our main estimates. Consistent with our expectation, we find the "bench sciences" such as most disciplines under life sciences and physical sciences — where experiments have to be conducted, coordinated, and troubleshot, and protocols and instruments adjusted, in real time — are also where the synchrony advantage is most robust. While neuroscience shows similar patterns, we consider psychology as an exception, given much of its experiments require face-to-face interaction with participants and are therefore constrained to local settings(21). Among social sciences, we find a qualitative-quantitative divide: whereas gender studies rely more on synchronous collaboration on interpretation and writing, demography and economics show no synchrony advantage given that authors spend more time analyzing data asynchronously (27). Economic and management science overall is also organized around a strong working paper culture that supports asynchronous exchange of ideas, which also explains the universal null effect (28). We argue that these findings suggest an interaction between videoconferencing facilitating a growing expectation for more synchrony in general in scientific collaboration and the temporal organization of scientific tasks in a particular field.

## The Synchrony Advantage Imposes Unequal Burden Across Countries

Because working-hour overlap is unevenly distributed by geography, a synchrony advantage also does not fall evenly across the world. Researchers in time zones far from the hubs of scientific activity have fewer collaborators with whom they can share working hours. To make the implications for global inequality in science more intuitive, we conduct a naive thought experiment: what if a country's international collaborators had all worked from its own time zone with maximum overlap in working hours? Using the estimated average gap of +2.8% publications per hour of overlap post-pandemic, we price the collaborative output each country forgoes, relative to this counterfactual, conditional on its existing international collaboration network. For better coverage of developing countries, we expanded our sample to include authors from the top 1,500 Nature Index institutions. For every country, we ask how many internationally collaborative publications it would have produced during 2021–2025 had its authors shared the same time zone with every one of their collaborators (Fig. 5, Panel A). Comparing the real growth against this counterfactual scenario, the world could have missed up to 11% potential international publications: Australia and New Zealand could have missed up to 16–17%, East and Southeast Asia 11–16%, Latin America 8.5–14%, and the Middle East 10–12%, against 6–10% across Europe.

These burdens hit harder where no synchronous alternative exists. We discount each country's burden by the availability of potential collaborators within an hour of its working day — weighting Panel A by the share of the world's pre-2020 scientific output produced outside of the country's time zone (Fig. 5, Panel B). The United States, despite bearing a large burden from low-overlap collaboration, sheds nearly a third of it as 30% of the world's pre-2020 output sits within the same time zone, whereas the cost is almost unavoidable for Oceania, South and West Asia, as well as Latin America. These estimates suggest that the rising expectation of synchronous collaboration might have imposed its steepest costs precisely on Scientific Global South countries that are in the middle of the global scientific hierarchy, whose working days barely intersect the world's centers of scientific activity — while Europe, embedded in the largest block of synchronized capacity, bore the least.

## Discussion

We found that from 2021 onward, authors and institution pairs that share more working hours develop more new collaborations, work with more collaborators, and publish more together than pairs with less overlap. This emerging synchrony advantage cannot be explained by other Covid-related shocks and is concentrated in fields organized around interdependent, real-time coordination. We argue that one possible explanation is that as synchronous interactions become both easier and expected, collaborators with more overlap in working hours will gain an advantage over those with less overlap because they face fewer scheduling constraints and are more likely to find a mutually convenient time to meet. These findings reveal how technological change can shape the temporal organization of science, with broader implications for remote work, social impacts of technology, social studies of science, and the future of funding policy and international collaboration.

First, time shapes science by structuring when and how remote work is possible. As remote work becomes more prevalent, we are tempted to assume that technology has made it easier for everyone. Science provides a reality check. Even among scientists with considerable autonomy over their schedules and collaborators, time remains a significant constraint. As remote work becomes increasingly transnational and spanning time zones, our findings highlight the need to understand how expectations for synchronous interactions may impose on the scientific community and beyond.

Second, time provides a unique angle through which to understand the complex and sometimes contradictory effect of technology on contemporary science. Science and technology studies have long argued that technological infrastructure is embedded within and co-evolves with existing organizational and social arrangement (29). The widespread application of videoconferencing during the pandemic, its benefits remained conditional by the temporal organization of scientific work. By introducing a new normative expectation that real-time coordination once impossible is now possible and preferable, digital technology made time differences, once declining in the age of emails, salient again and turned it into a new structural determinant of who gets to participate in science at a lower cost.

Third, digital technologies have made scientific work increasingly portable across space, but not all scientific work is equally portable across time. Disciplines differ in their temporal organization: the degree to which collaborative work requires simultaneous availability. As synchronous virtual interaction becomes more common, fields where work can be organized asynchronously may be better able to sustain geographically dispersed collaborations, while fields organized around a high level of synchronicity may remain more constrained by time differences. The field heterogeneity we find offers some clarification on the scope of this mechanism: disciplines where collaboration instead depends on shared physical infrastructure (e.g., large-scale instrumentation in high-energy physics and astronomy) or in-person fieldwork (e.g., ecology) are less to exhibit the same synchronicity advantage. The "death of distance" happens differently across disciplines depending on the extent to which their core research activities can be moved online. Future research should examine how these differences shape not only the extent and output, but also the heterogeneity of scientific work.

Finally, as collaboration becomes increasingly organized around synchronous virtual interactions, institutions that share working hours with global hubs gain a structural advantage. Scientific resources like funding, prestige, and collaboration opportunities are varied across location and unevenly distributed, institutions are different systematically to the extent they share overlapped working hours with major research hubs. As collaboration becomes increasingly organized around an expectation of synchronous work, individuals and organizations whose working hours align better with global research hubs will enjoy a structural advantage in forming and sustaining a transnational workflow. The expansion of Zoom does not flatten the unequal geography of science but reconfigures it—making time a new axis of inequality in knowledge production. For policy makers, our findings suggest that the institutionalization of synchronous virtual collaboration has aggravated existing geographical hierarchies in science. Governments, funding agencies, and professional associations that wish to promote scientific collaborations around the world should consider how to redistribute resources, facilitate travel, and build infrastructure that supports teams working across time zones.

## Materials and Methods

**Data Quality**. We exclude within-country collaborations. We also exclude publications that have more than 30 authors and fields that have distinct collaboration structures, including high-energy physics and astronomy, where collaboration is structured around large-scale instruments, and medical sciences, where collaboration was accelerated by a shift toward COVID-19 research. Mathematics and computer science are also not included in the sample due to their atypical publication cycles compared to other fields.

To assess the reliability of OpenAlex data underlying our variables, we draw a stratified random sample of publications by scientific field and publication year and manually validate each author's affiliation information. Among 654 papers we sampled, OpenAlex records correctly identify 94% of the international collaborations between institutions, and 89.1% between authors, comparable to the accuracy rates of 93% reported in prior validations of large-scale bibliometric databases by OpenAlex. Given this degree of accuracy, we do not expect misclassification to systematically bias our estimates.

**Model Specification**. We estimate difference-in-differences models at two levels of analysis: an author × collaborator-time-zone panel and an institutional-pair panel.

The author-level analysis captures how an individual researcher's collaboration with researchers in different time zones changes over time. For author *a*, time zone *z*, and year *t*, the outcomes are (1) whether the author forms a first-time collaboration with a coauthor in that time zone, (2) the number of distinct collaborators in time zone *z*, and (3) the number of papers coauthored with researchers in that time zone. $Overlap_{a,z}$ is the number of shared working hours between the author's and the collaborator's time zones, capped at 7 hours. We estimate:

$$Y_{a,z,t} = \beta(Overlap_{a,z} \times Post_t) + \gamma Overlap_{a,z} + \alpha_a + \delta_{z,t} + \varepsilon_{a,z,t} \qquad \textbf{[1]}$$

where $Post_t$ equals 1 for 2021–2025 and 0 for 2015–2019, α denotes author fixed effects, and δ denotes collaborator-time-zone × year fixed effects, which absorb time-invariant heterogeneity and any time-variant shock to the pool of collaborators in a given time zone. The coefficient $\beta$ captures the change after 2020 in the association between working-hour overlap and collaboration outcomes, comparing how an author's collaboration with higher-overlap time zones changes after 2020 relative to the same author's collaboration with lower-overlap time zones.

The institutional-pair analysis captures changes in collaboration between pairs of institutions *i* and *j* in different countries. The outcomes are (1) the number of distinct collaborating researcher pairs in year *t* and, (2) the number of papers coauthored by researchers at the two institutions. We estimate:

$$Y_{i,j,t} = \beta(Overlap_{i,j} \times Post_t) + \alpha_{i,j} + \delta_{i,t} + \delta_{j,t} + \varepsilon_{i,j,t} \qquad \textbf{[2]}$$

where $\alpha_{i,j}$ denotes institution-pair fixed effects, and $\delta_{i,t}$ and $\delta_{j,t}$ denote institution × year fixed effects for each member of the pair. Because overlap is time-invariant within a pair, its main effect is absorbed by $\alpha_{i,j}$. Relative to Eq. **1**, this design absorbs time-variant shocks to both institutions, such as changes in either institution's size, funding, or output. We assume, conditional on these fixed effects, no pair-specific shock such as bilateral travel restrictions that correlates with the pair's working-hour overlap.

We estimate Eq. 1 using ordinary least squares as a linear probability model for the binary outcome, and Eqs. 1 and 2 by Poisson pseudo-maximum likelihood (PPML) for the count outcomes. The PPML specification assumes that, absent the pandemic, high- and low-overlap pairs would have experienced equal proportional rather than equal absolute changes (30). We consider this a more plausible counterfactual because collaboration tends to grow multiplicatively. Standard errors are clustered by time-zone pair in both designs.

**Acknowledgments**

We are grateful to Ronald Breiger, Likun Cao, Joseph Galaskiewicz, Charles Gomez, Erin Leahey, Diego Leal, Jina Lee, Daniel Menchik, Brendan Tinoco, Mo Xiao, Yulin Yu, and the audience at ICSSI 2026 for their insightful and helpful feedback. We thank members of the Medical Sociology Shop at the University of Arizona (Lucia Alfano, Katrina Market, Zeke Pangelinan, Leeah Payne, Jonathan Reyes, and Mariandrea Valenzuela) for validating OpenAlex data.

## Figures and Tables

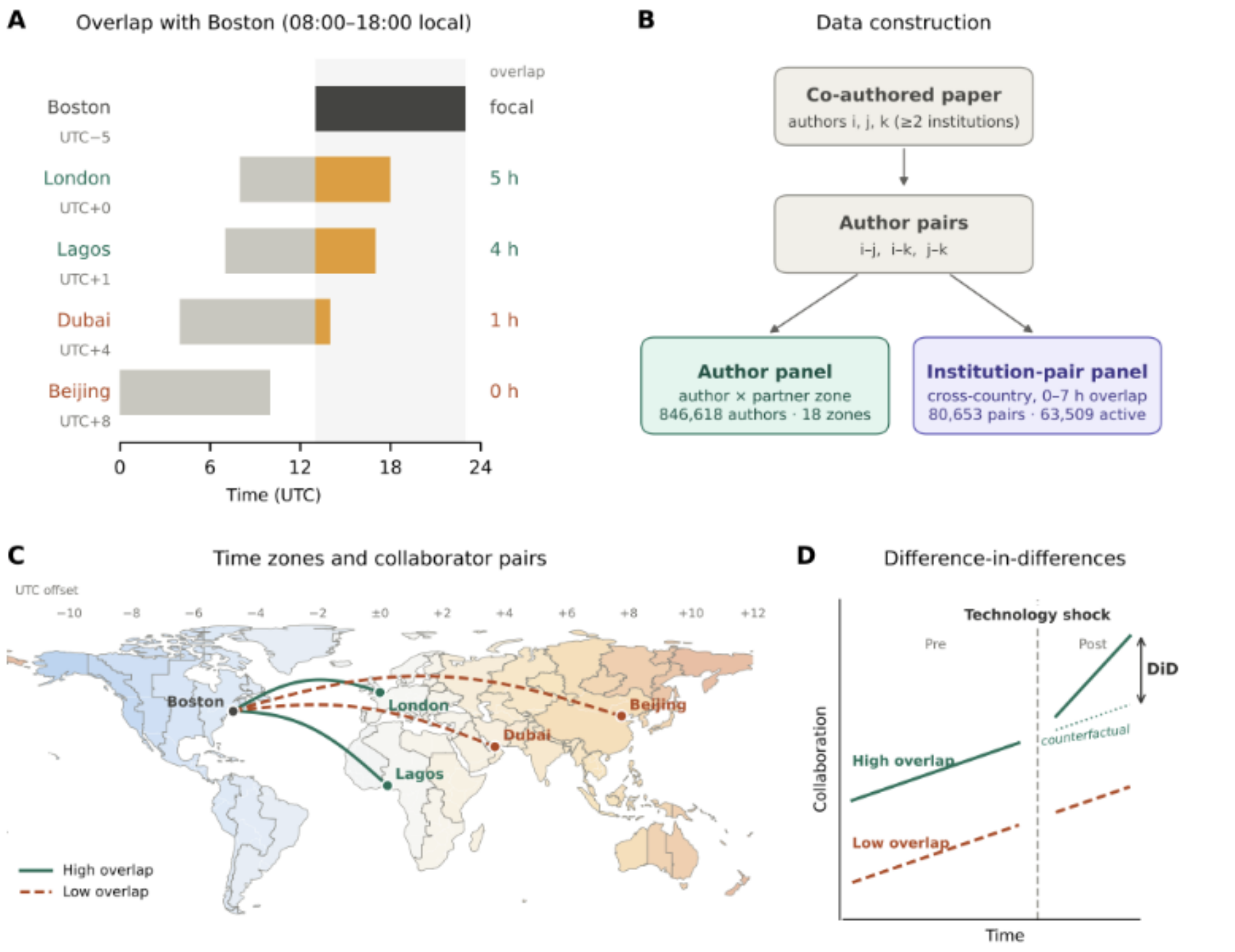

**Figure 1.** We exploit the COVID-induced technological shock and working-hour overlap between time zones to trace the emergence of "synchrony advantage" in a difference-in-differences design. Panel A shows working-hour overlap between and four cities, defined as the number of hours overlapped during 8am-6pm, orange segments indicate hours shared with Boston. Panel B demonstrates the construction of two analytical panels. Each cross-institution paper is decomposed into author pairs, which are aggregated into an author level and an institutional panel. Panel C shows the example pairs for high overlap and low overlap working-hours. Panel D illustrates how the difference-in-differences design uncovers the synchrony advantage. We compare collaboration between high- and low-overlap pairs before and after technology shock. The dotted line shows the counterfactual trajectory of high-overlap pairs under parallel trends; the vertical gap is the difference-in-differences estimate.

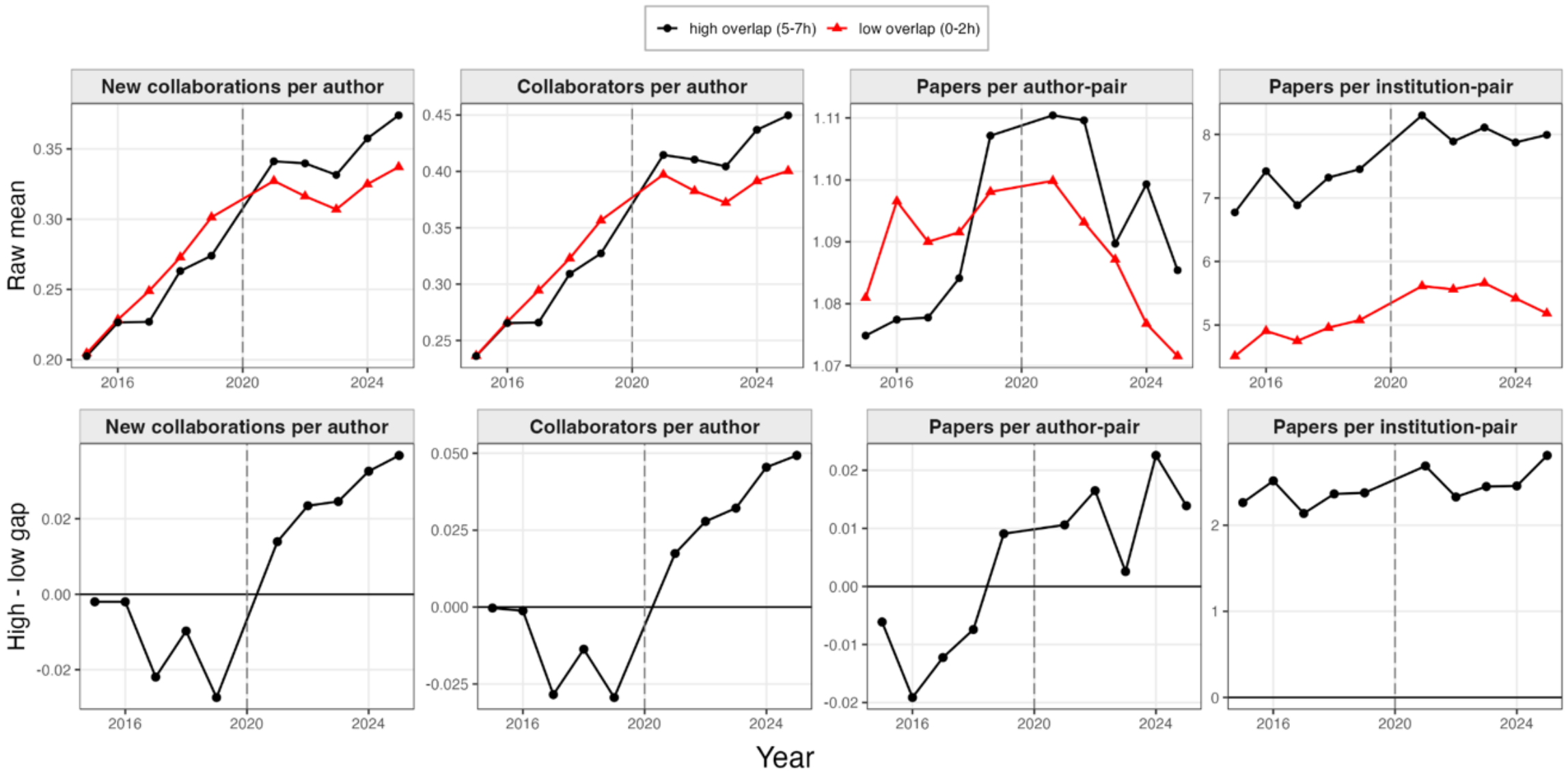

**Figure 2.** International collaboration by working-hour overlap, 2015–2025. Yearly means for collaborations with high (5–7 h) and low (0–2 h) working-hour overlap: (A) new collaborations per author, (B) collaborators per author, (C) papers per author pair, and (D) papers per institution pair. (Top) Levels; (Bottom) the high-minus-low difference.

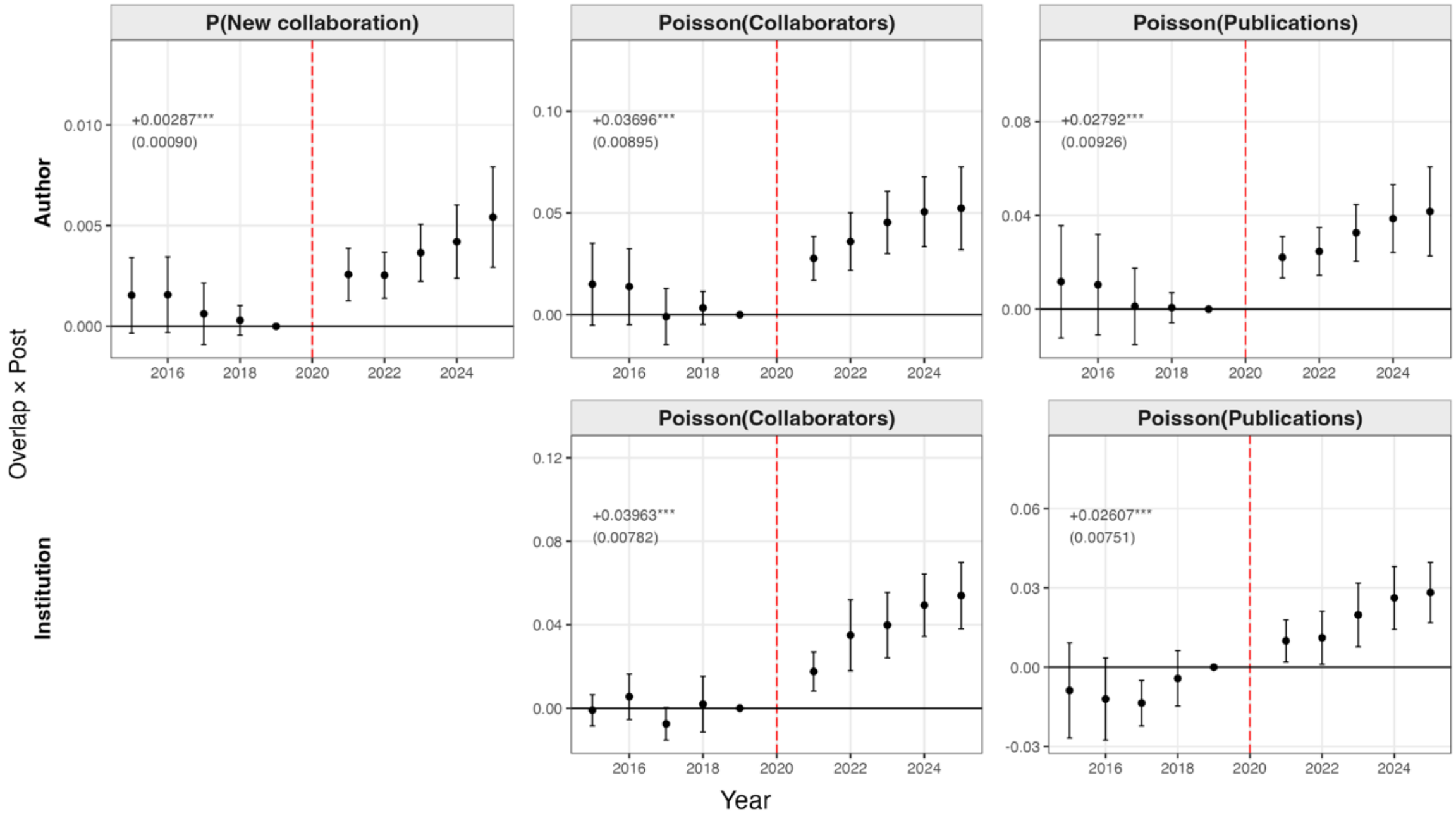

**Figure 3.** The synchrony advantage widens after 2020 across all five outcomes. Event-study estimates of Overlap × Year (reference 2019; 2020 omitted) with continuous working-hour overlap (0–7 h). In-panel labels give the pooled Overlap × Post estimate. Error bars are 95% confidence intervals from standard errors clustered by time-zone pair. ***P < 0.01, **P < 0.05, *P < 0.1.

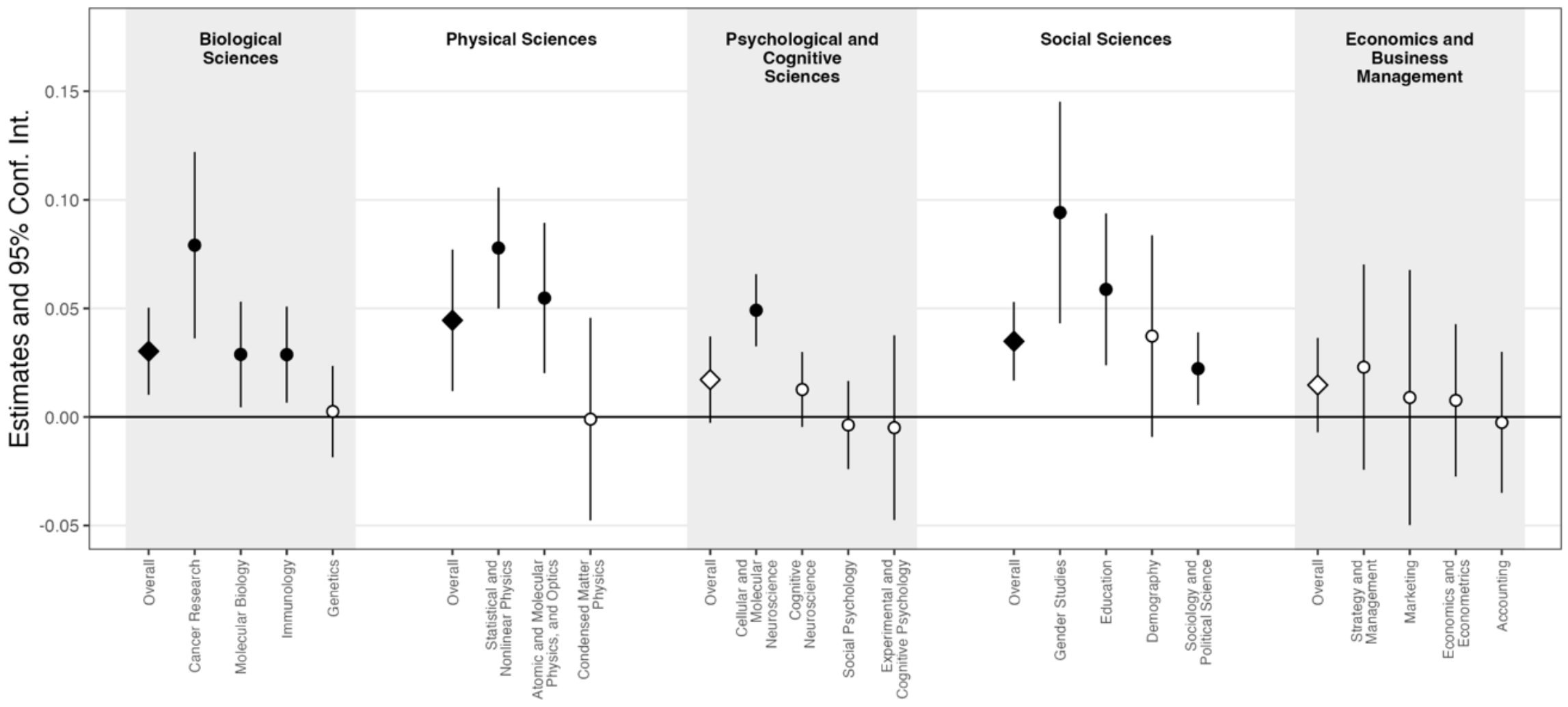

**Figure 4.** The synchrony advantage can be found across domains but unevenly distributed across fields and absent in experimental psychology and economics. Overlap × Post estimates for Poisson(Publications) in the author design, by research domain and subfield. Filled markers denote $P < 0.05$ under a bootstrapped null; horizontal bars are 95% confidence intervals. All three outcomes appear by discipline in Table 2.

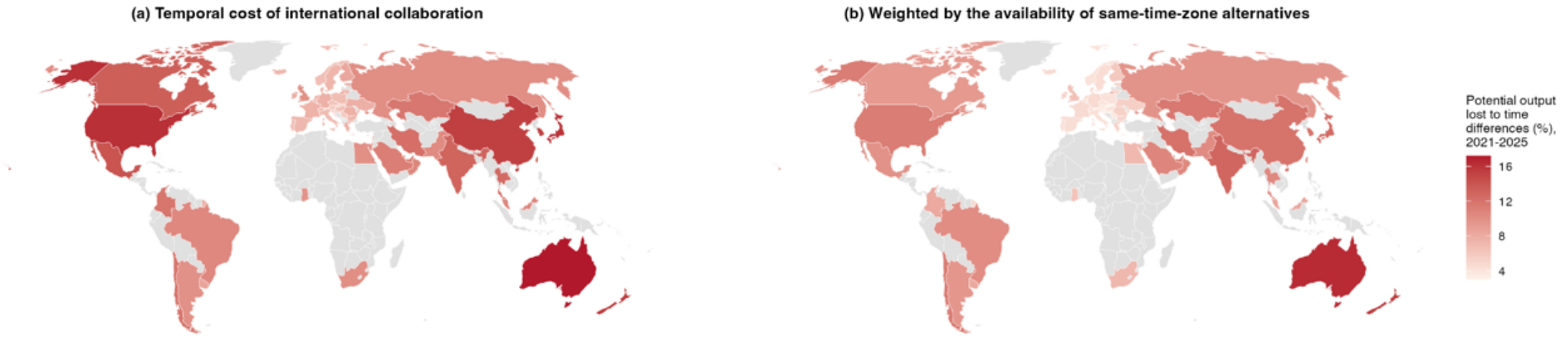

**Figure 5.** The temporal burden of international collaboration. (A) For each country, one minus the ratio of actual to counterfactual growth in international collaborative publications in 2021–2025, where the counterfactual uses the estimated average gap of +2.8% publications per hour of overlap post-pandemic to estimate what if there was no time difference between any collaborators. (B) The same counterfactual weighted by the share of the world's pre-2020 research output located within one hour of the country's time zone.

| | (1) | (2) | (3) | (4) | (5) |
|---|---|---|---|---|---|
| | Author | | | Institution | |
| | New collab. | Collabo-rators | Publica-tions | Collabo-rators | Publica-tions |
| Overlap × Post | 0.0029 | 0.0370 | 0.0279 | 0.0396 | 0.0261 |
| | (0.0009) | (0.0090) | (0.0093) | (0.0078) | (0.0075) |
| Pretreatment avg. | 0.107 | 0.300 | 0.146 | 3.783 | 2.416 |
| $R^2$ | 0.038 | 0.191 | 0.111 | 0.635 | 0.691 |
| No. of obs. | 13,876,650 | 13,876,650 | 13,876,650 | 577,510 | 633,108 |
| Author FE | Y | Y | Y | | |
| Time Zone × Year FE | Y | Y | Y | | |
| Institution Pair FE | | | | Y | Y |
| Institution × Year FE | | | | Y | Y |
| Poisson | | Y | Y | Y | Y |

**Table 1.** Difference-in-differences estimates: the gap in international collaboration by working-hour overlap widened after 2020.

PNAS

## Supporting Information for

# Technology Creates a New Synchrony Advantage in Global Scientific Collaboration.

Zheng Fu[a,1,2] and Zhuofan Li[b,1,2]

[a]School of Sociology, University of Arizona

[b]Department of Sociology, Virginia Tech

[1]To whom correspondence may be addressed.

**Email:** zfu@arizona.edu; zhuofan@vt.edu.

**This PDF file includes:**

Figures S1 to S13
Tables S0 to S13

**Other supporting materials for this manuscript include the following:**

## Figures

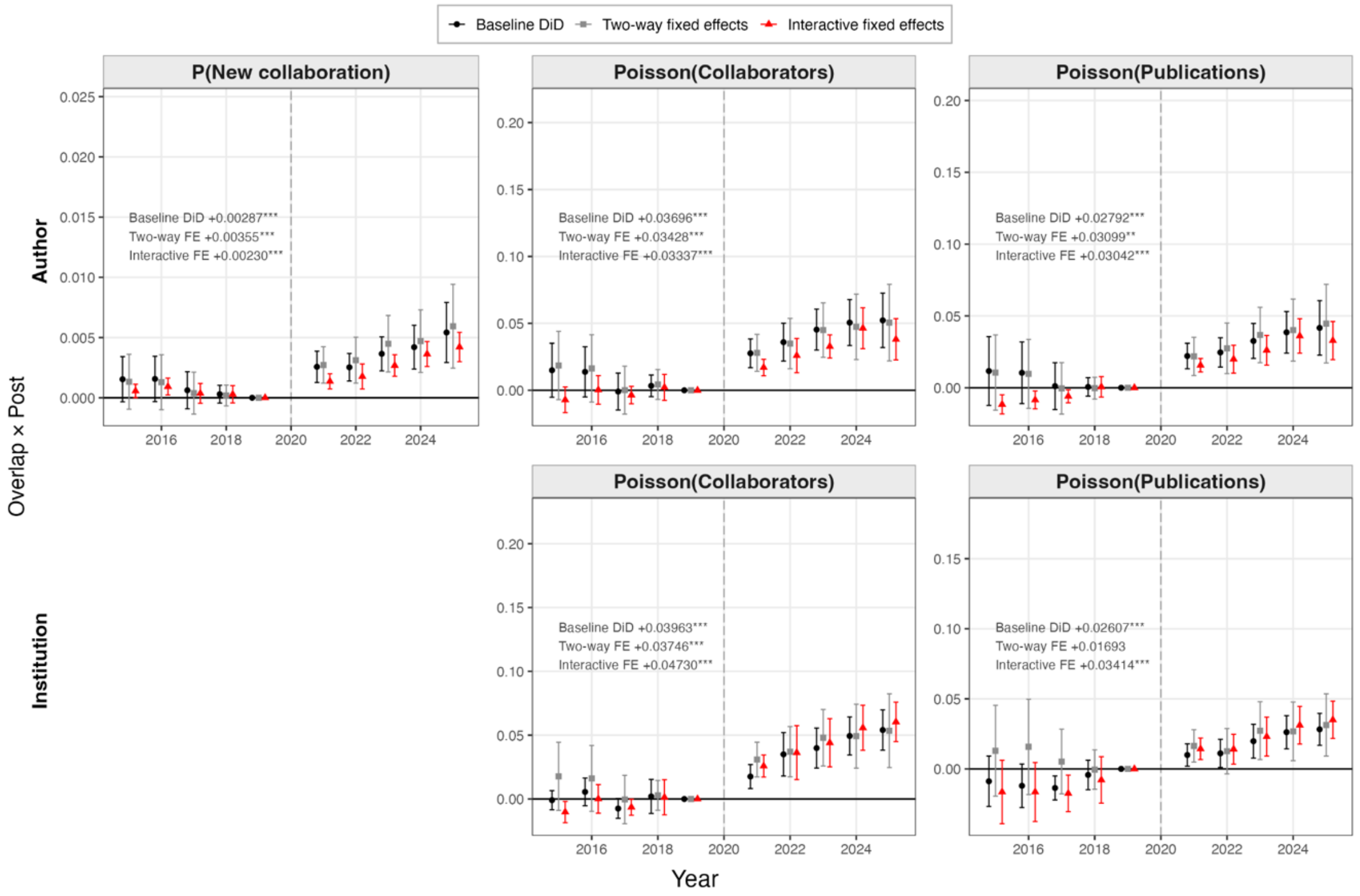

**Fig. S1. The main event studies are robust to alternative fixed effects.** Overlap × Year coefficients (reference 2019) under the baseline, two-way (Table S1, Panel A), and interactive (Table S1, Panel B) specifications. Error bars represent 95% confidence intervals.

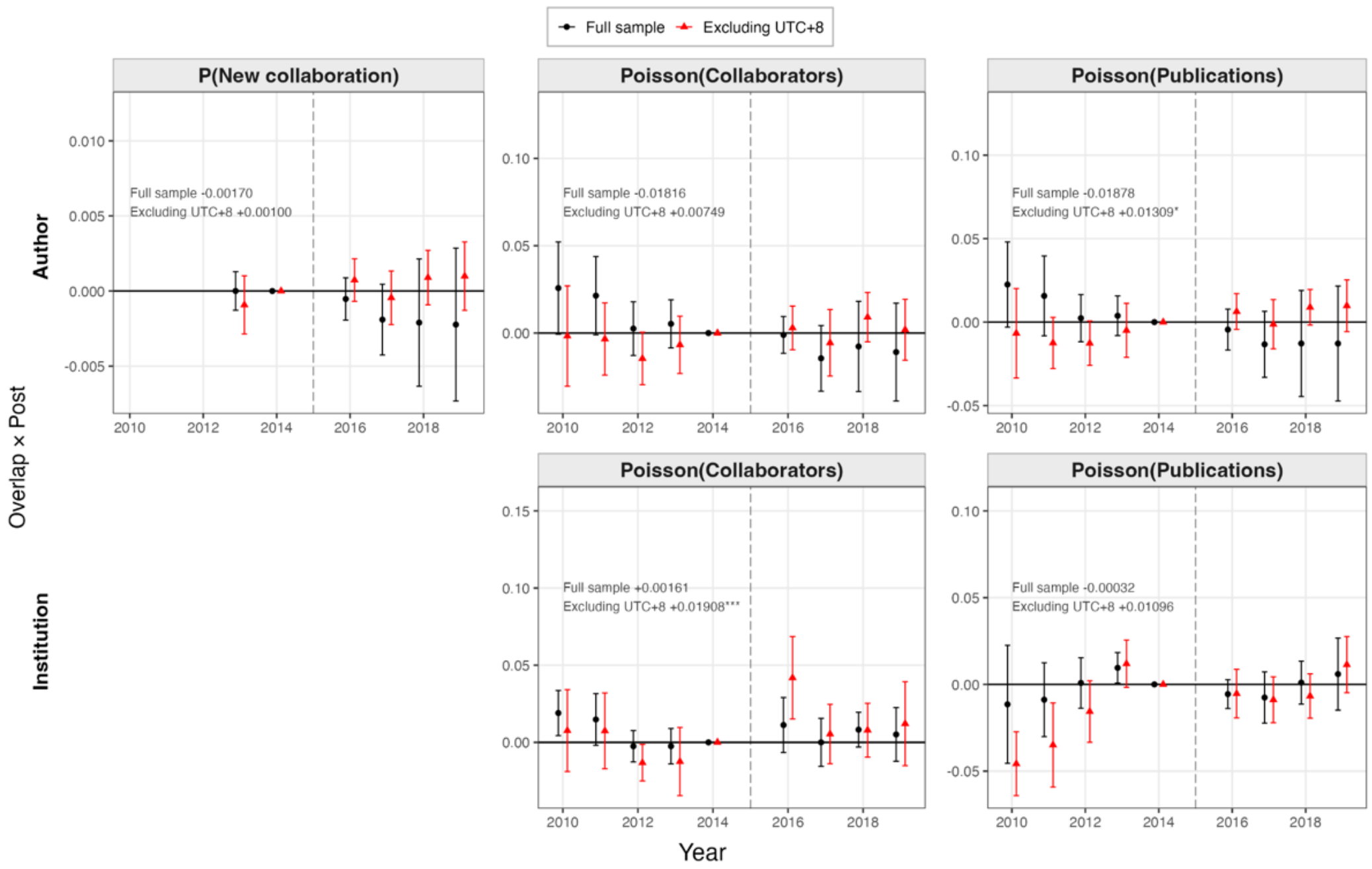

**Fig. S2. Placebo Test for full sample and Excluding UTC+8 results.**

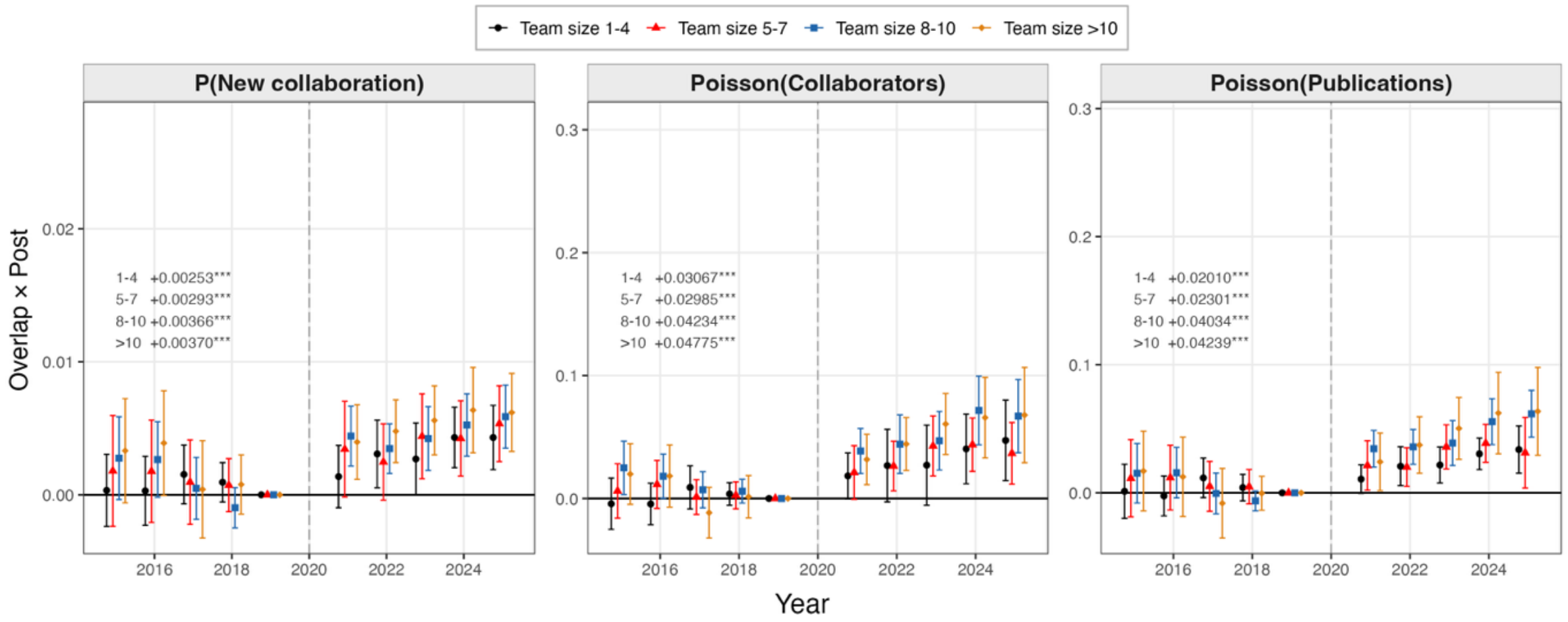

**Fig. S3. "Synchronicity advantage" by pre-pandemic team size**. Event-study estimates of the effect of working-hour overlap on collaboration, estimated separately by authors' median team size in 2015–2019: 1–4 (black circles), 5–7 (red triangles), 8–10 (blue squares), and more than 10 coauthors (orange diamonds). Author × partner time zone panel. Coefficients are relative to 2019; 2020 is omitted. Bars indicate 95% confidence intervals. Annotations report pooled Overlap × Post estimates for each group (***$P < 0.001$). Standard errors are clustered by time-zone pair.

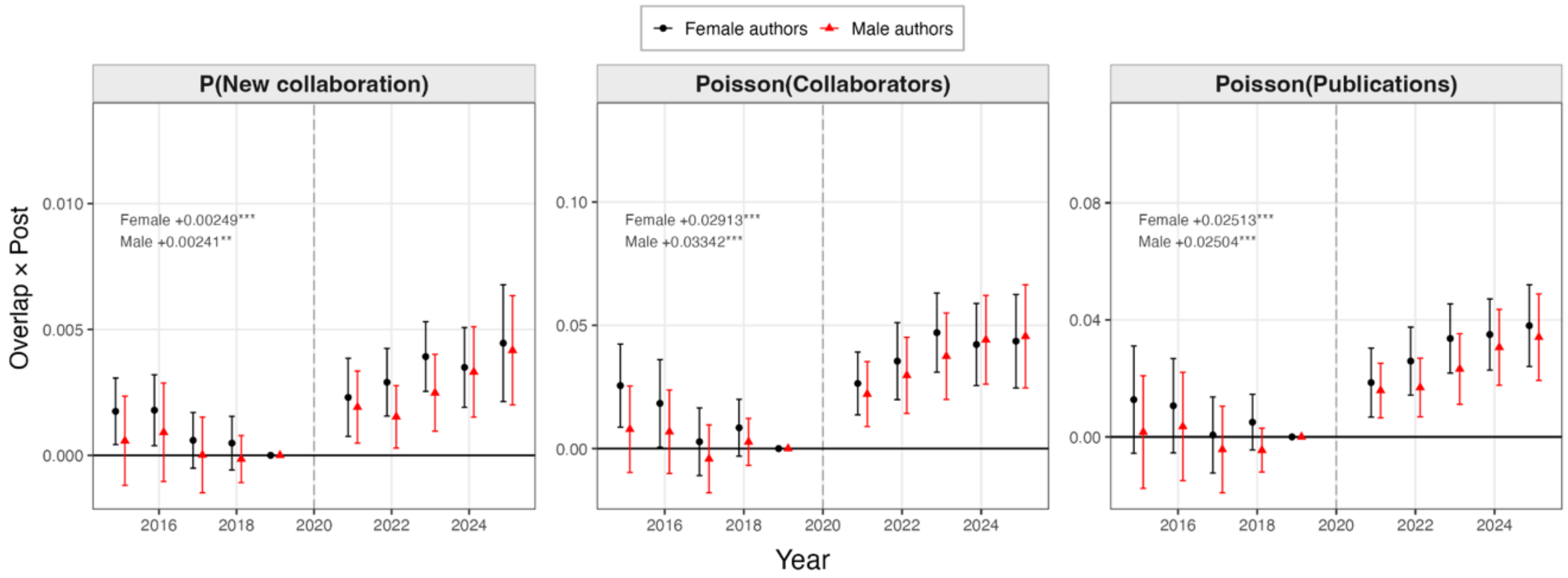

**Fig. S4. "Synchronicity advantage" by author gender.** Estimates of the effect of overlapping working hours on collaboration, based on an event-study design, are presented separately for female authors (black circles) and male authors (red triangles). Coefficients are normalized relative to 2019; data from 2020 are excluded from the model. Error bars represent 95% confidence intervals. Annotations indicate the aggregated "Overlap × Post" estimates for each group (**P < 0.01, ***P < 0.001). Gender is inferred from names, these results remain robust when using different classification thresholds or excluding institutions in East Asian.

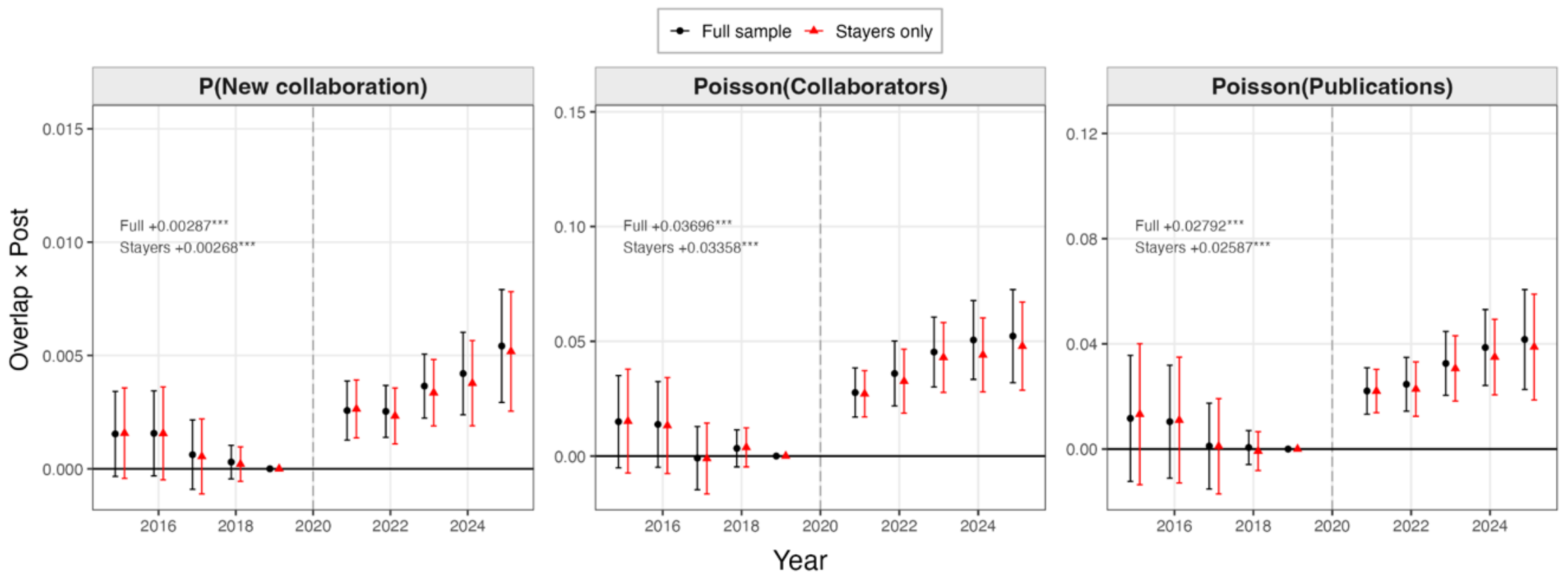

**Fig. S5. Synchrony advantage among authors who never changed time zones:** Event-study estimates of the effect of working-hour overlap on collaboration for the full sample (black circles) and for authors whose home time zone did not change over 2015–2025 (red triangles; 81% of authors). Author × partner time zone panel. Coefficients are relative to 2019; 2020 is omitted. Bars indicate 95% confidence intervals. Annotations report pooled Overlap × Post estimates for each group (***$P < 0.001$). Standard errors are clustered by time-zone pair.

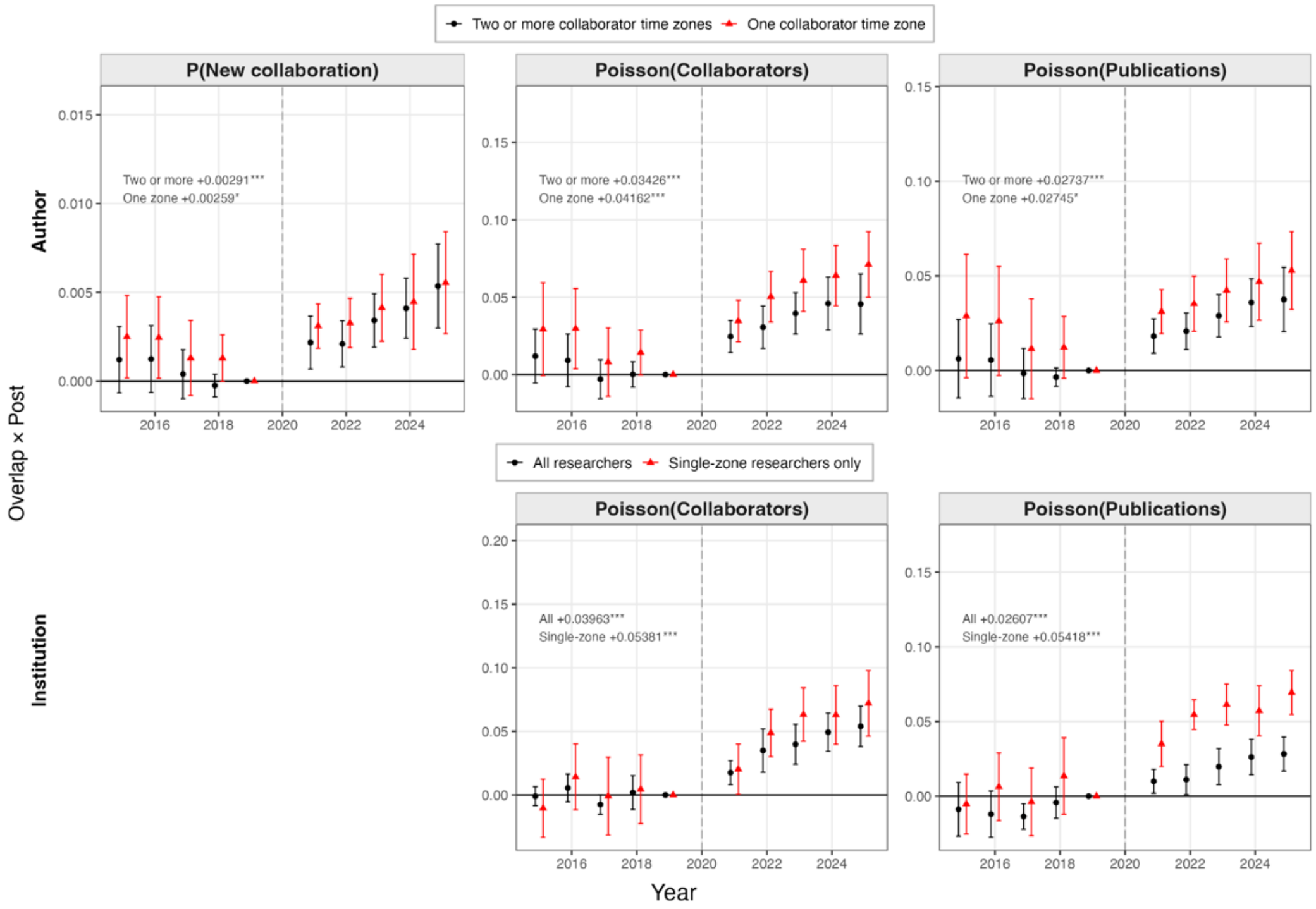

**Fig. S6. The "synchronicity advantage" in collaborations with partners in the same time zone.** The impact of overlapping working hours on collaboration is estimated using an event-study design. Top panel is author–partner time zone combinations; estimates are presented for two groups of authors—those with collaborators spread across two or more time zones (black circles) and those whose collaborators are all located in the same time zone (red triangles). Bottom panel is institution-pair combinations; estimates are presented for all researchers (black circles) and for researchers whose collaborators are all in the same time zone (red triangles). Regression coefficients are normalized relative to 2019; data from 2020 are excluded from the model. Error bars represent 95% confidence intervals. Labels indicate the aggregated estimates for the "Overlap × Post" interaction term for each group (*P < 0.05, ***P < 0.001). Standard errors are clustered at the "time zone pair" level.

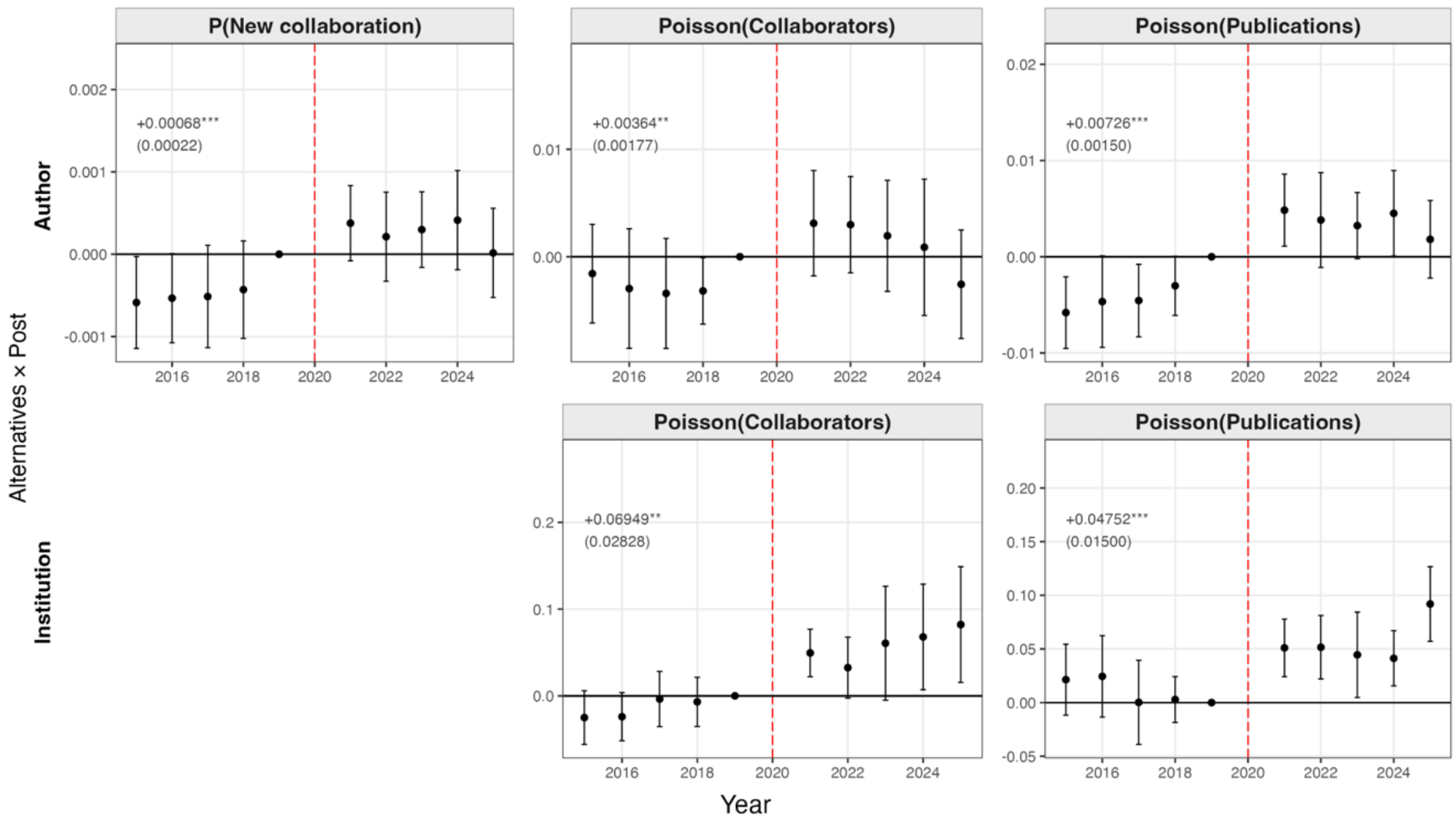

**Fig. S7. Post-2020 effects of alternative collaborators.** Top: author × partner time zone panel. Bottom: institution-pair panel; new collaboration is not defined at the institution level. Coefficients are relative to 2019; 2020 is omitted. Bars indicate 95% confidence intervals. Annotations report pooled Alternatives × Post estimates with standard errors in parentheses (**P < 0.01, ***P < 0.001). Standard errors are clustered by time-zone pair.

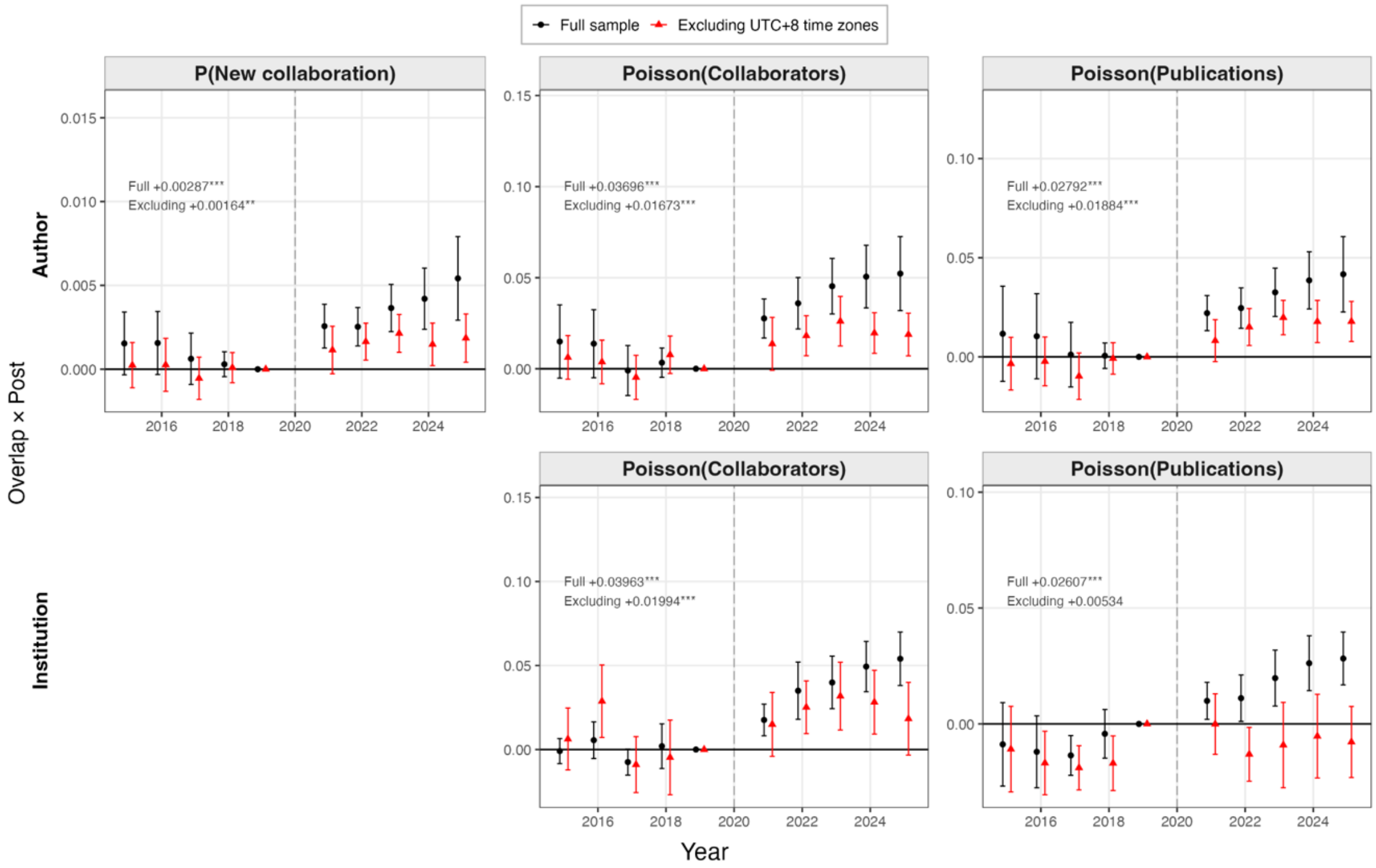

**Fig. S8. Synchrony advantage excluding UTC+8 time zones.** Event-study estimates of the effect of working-hour overlap on collaboration for the full sample (black circles) and excluding all collaborations involving institutions in UTC+8 (red triangles), including mainland China, Hong Kong, Taiwan, Singapore, and Malaysia…Top panel is author × partner time zone panel, and bottom is institution-pair panel; new collaboration is not defined at the institution level. Coefficients are relative to 2019; 2020 is omitted. Annotations report pooled Overlap × Post estimates for each group (**P < 0.01, ***P < 0.001). Standard errors are clustered by time-zone pair.

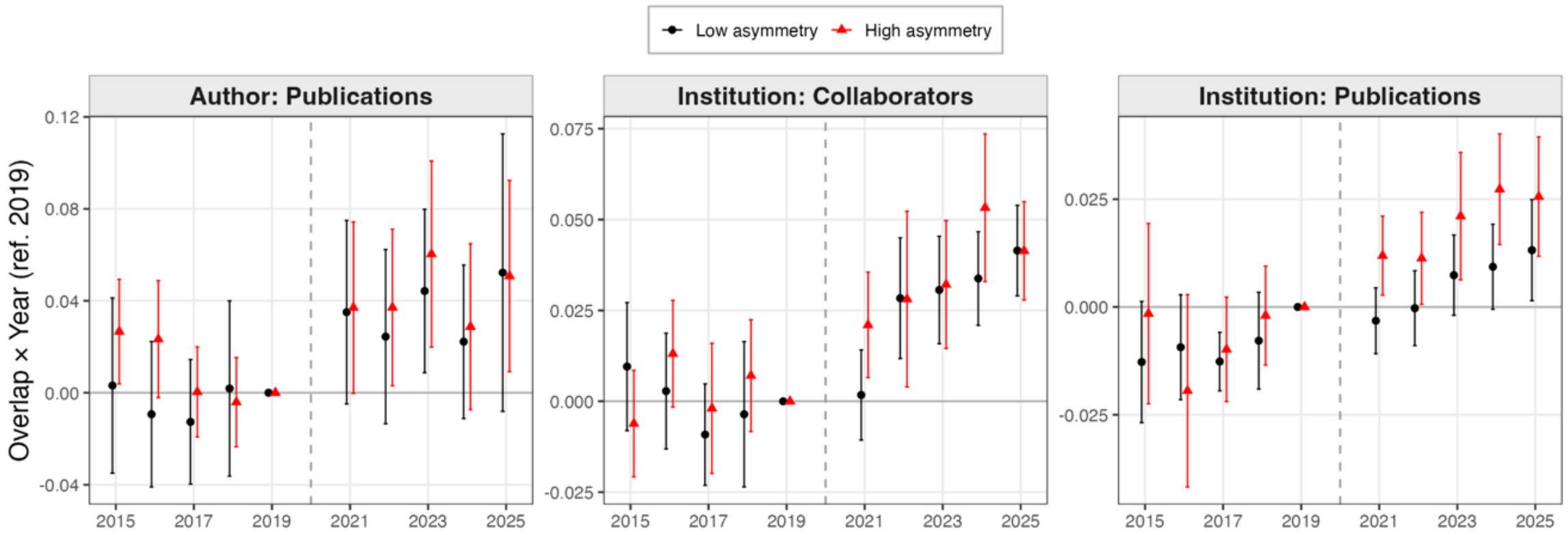

**Fig. S9. Synchrony advantage by status asymmetry between collaborators.** Each data point displays the estimated coefficient for the "Overlap × year" term at varying levels of "asymmetry in intra-group status."

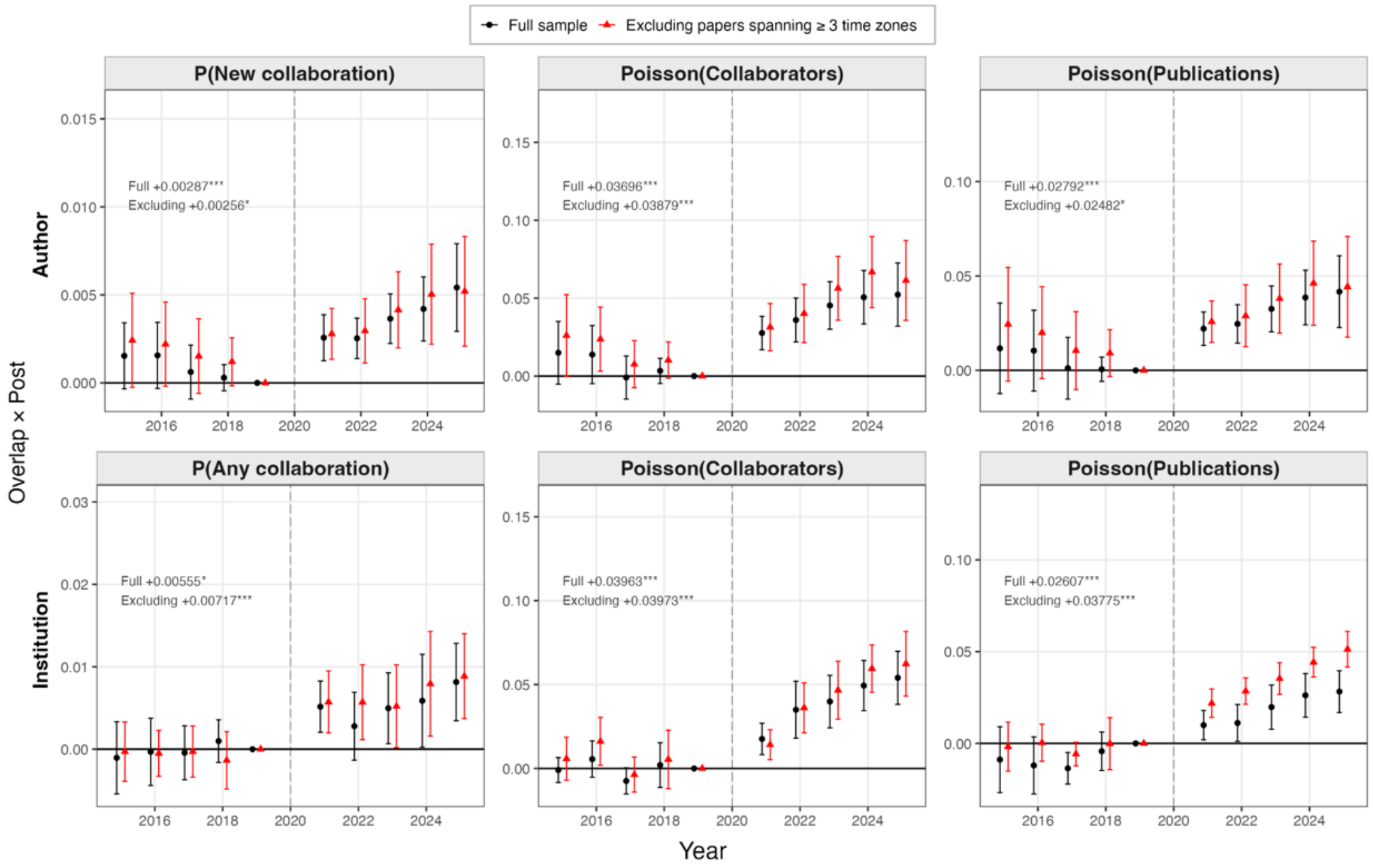

**Fig. S10. Excluding collaborations that span 3 or more time zones.** Event-study estimates of the effect of working-hour overlap on collaboration for the full sample (black circles) and excluding all works that span three or more distinct time zones (red triangles) — works whose authors do so in the author design, and whose institutions do so in the institution-pair design — with first-collaboration histories recomputed on the restricted corpus. Top panel is the author × partner time zone panel; bottom is the institution-pair panel; new collaboration is not defined at the institution level. Coefficients are relative to 2019; 2020 is omitted. Error bars represent 95% confidence intervals. Annotations report pooled Overlap × Post estimates for each group (*P < 0.05, **P < 0.01, ***P < 0.001). Standard errors are clustered by time-zone pair.

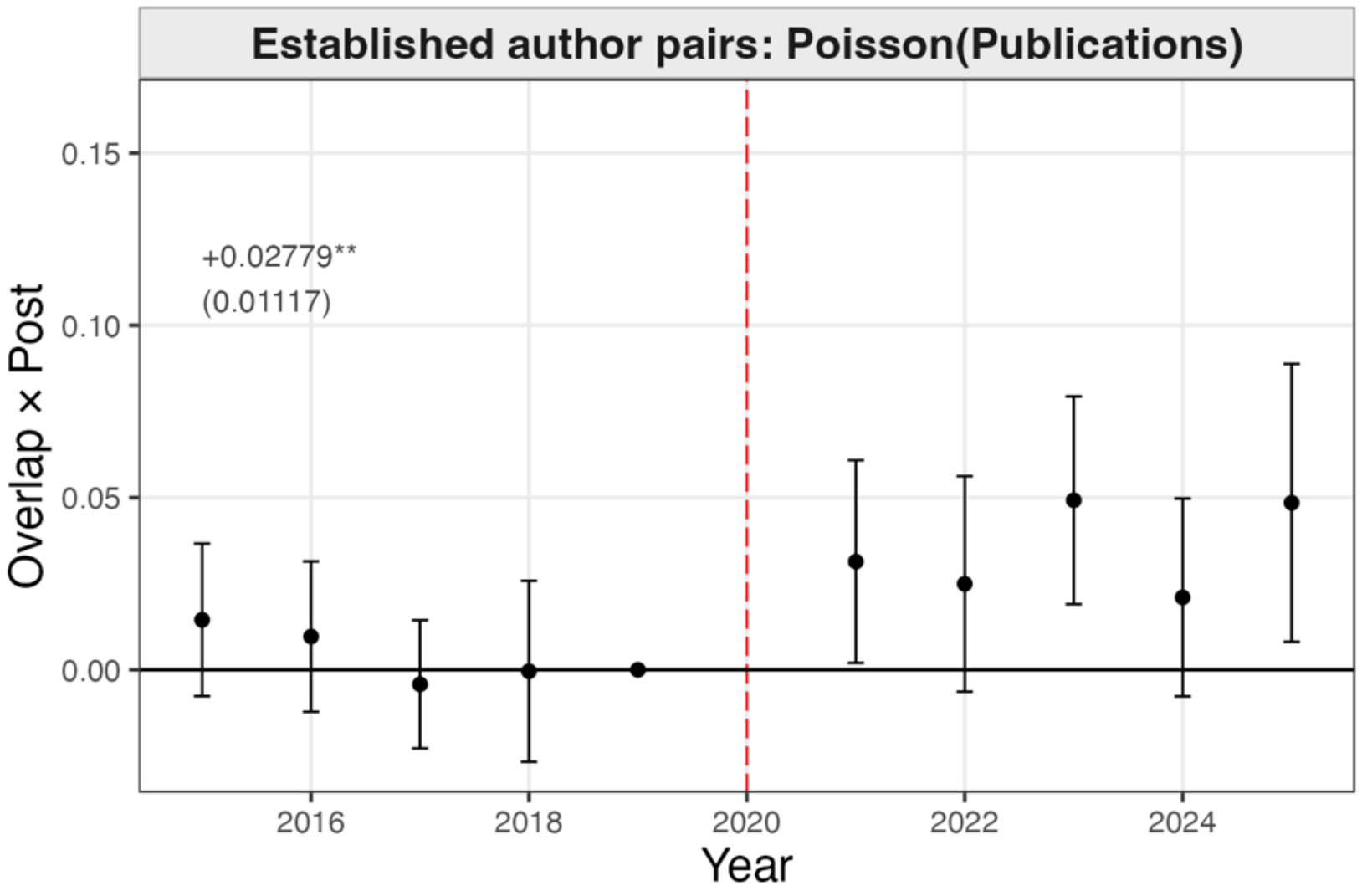

**Fig. S11: Synchrony advantage among established author pairs.** Overlap × Year event-study estimates from a Poisson regression of the number of coauthored publications, on author pairs who had collaborated between 2010–2014, with pair and year fixed effects. We find that even pre-existing author pairs experienced a synchrony advantage in productivity only after 2020. Error bars represent 95% confidence intervals.

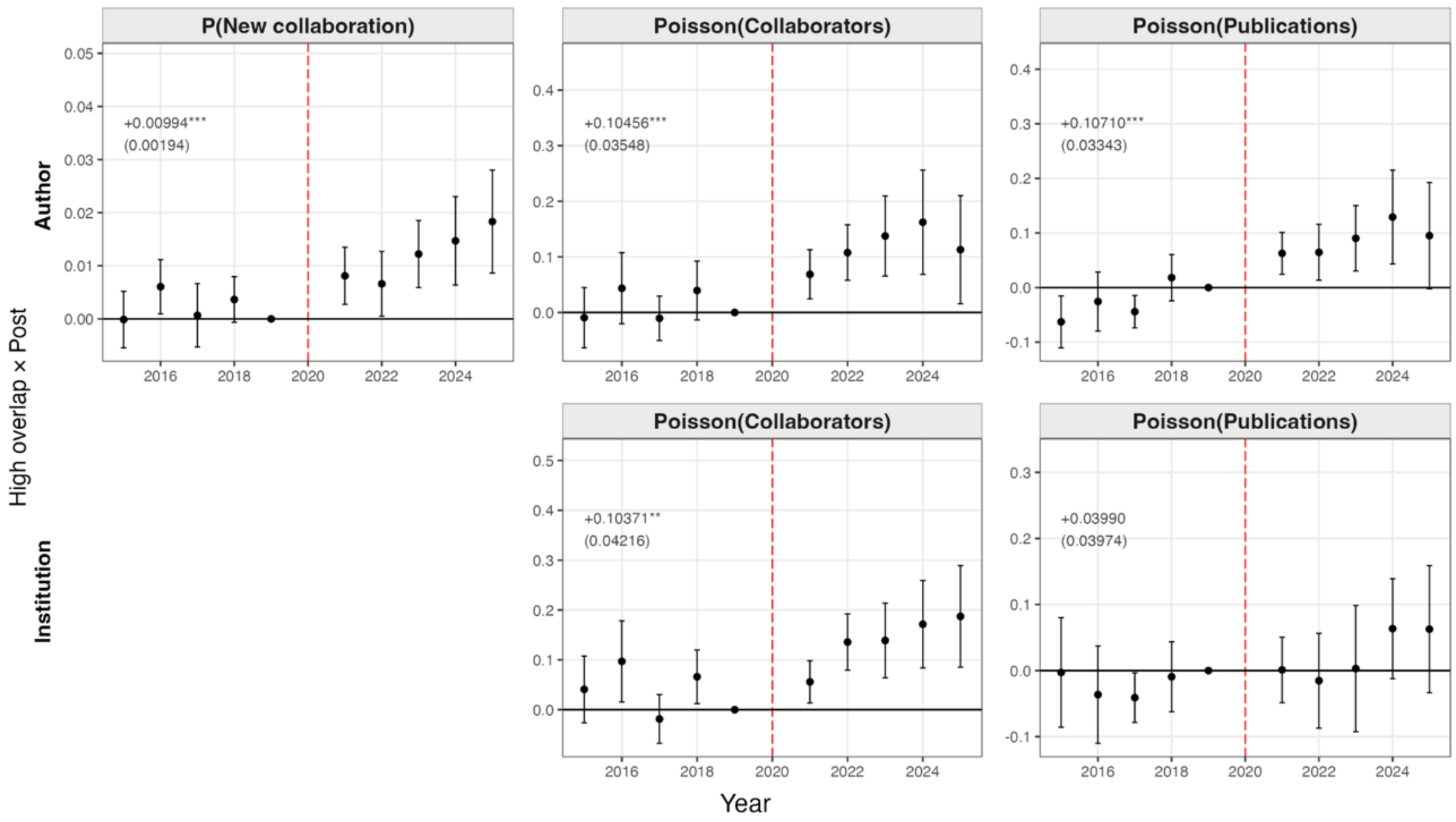

**Fig. S12. Binary treatment.** To relax the assumption that the synchrony advantage increases linearly in overlap, continuous overlap is replaced with a binary contrast between high-overlap (5–7 shared hours) and low-overlap pairs (0–2 hours; reference), excluding pairs with intermediate overlap (3–4 hours). Panels show High overlap × Year event-study estimates (reference: 2019; 2020 omitted) for the author design (top row; author × year fixed effects) and the institution-pair design (bottom row; new collaboration is not defined at the institution level). Annotations report pooled High overlap × Post estimates with standard errors in parentheses (*P < 0.05, **P < 0.01, ***P < 0.001). Error bars represent 95% confidence intervals. Standard errors are clustered by time-zone pair.

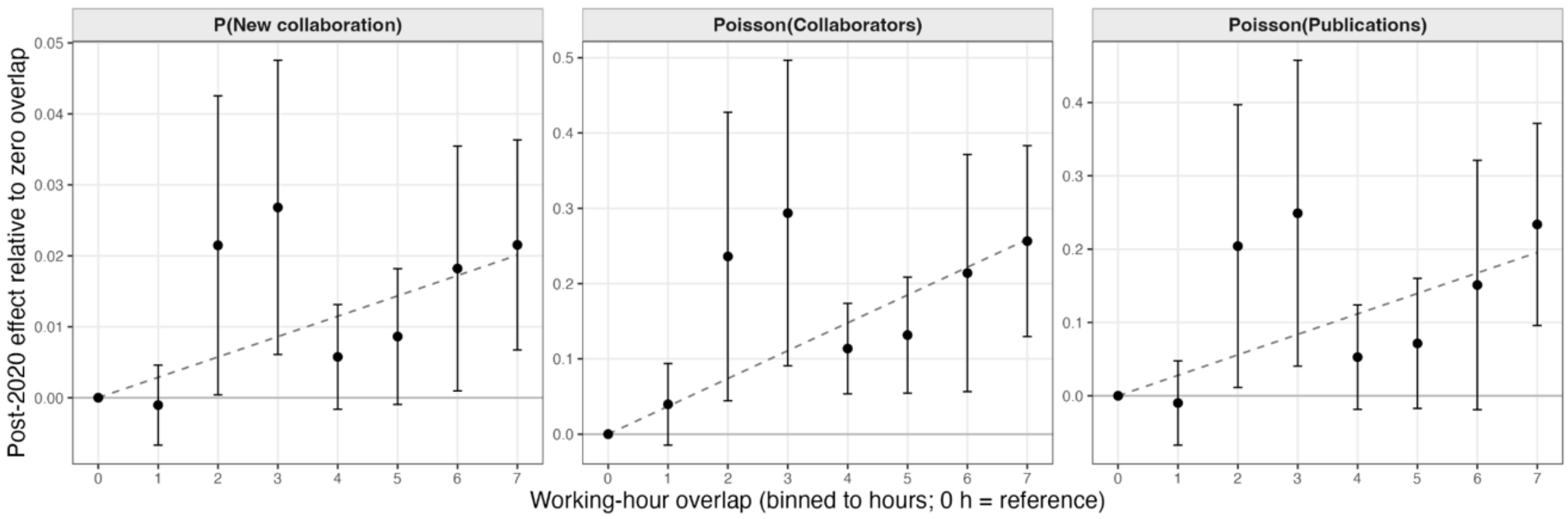

**Fig. S13. By-hour decomposition:** This figure relaxes the linearity assumption of the continuous specification by replacing Overlap × Post with a full set of overlap-hour dummies interacted with Post. Each point shows the post-2020 change in the outcome for author × collaborator-time-zone pairs at a given overlap, relative to pairs with no working hours overlap. One hour of overlap has no detectable effect on any outcome, consistent with synchronous collaboration requiring a minimum usable window in which a meeting can be scheduled. After two hours, the effect generally increases with overlap, and the estimates at seven hours closely match the linear prediction, indicating that the continuous estimate is not driven by any single overlap level.

## Tables

| | Low overlap (0–2 h) | | High overlap (5–7 h) | |
|---|---|---|---|---|
| | 2015–2019 | 2021–2025 | 2015–2019 | 2021–2025 |
| *Author* | | | | |
| New collaboration (0/1) | 0.108 (0.310) | 0.142 (0.349) | 0.102 (0.303) | 0.151 (0.358) |
| Collaborators | 0.295 (1.631) | 0.389 (2.760) | 0.281 (1.251) | 0.423 (1.763) |
| Publications | 0.146 (0.684) | 0.196 (1.456) | 0.138 (0.474) | 0.213 (0.774) |
| No. of obs. | 7,937,950 | | 1,797,860 | |
| *Institution-pair panel* | | | | |
| Collaborators | 2.995 (9.960) | 3.910 (12.048) | 4.253 (14.095) | 6.409 (19.000) |
| Publications | 1.922 (5.068) | 2.693 (6.642) | 3.361 (8.365) | 4.768 (10.652) |
| No. of obs. | 416,350 | | 61,300 | |

**Table S0. Summary Statistics.** Panels A and B report means (standard deviations) of each outcome for low-overlap (0–2 h) and high-overlap (5–7 h) pairs before and after the pandemic. Panel C describes the institution-pair risk set: each group's three leading country pairs and the share involving UTC+8 institutions, as a percentage of the group's joint publications and of its institution pairs.

*Panel A. Two-Way Fixed Effects*

| | (1) | (2) | (3) | (4) | (5) |
|---|---|---|---|---|---|
| | Author | | | Institution | |
| | New collab. | Collabo-rators | Publica-tions | Collabo-rators | Publica-tions |
| Overlap × Post | 0.0036 | 0.0343 | 0.0310 | 0.0375 | 0.0169 |
| | (0.0013) | (0.0133) | (0.0127) | (0.0136) | (0.0136) |
| Pretreatment avg. | 0.107 | 0.300 | 0.146 | 3.765 | 2.405 |
| $R^2$ | 0.037 | 0.190 | 0.111 | 0.616 | 0.668 |
| No. of obs. | 13,876,650 | 13,876,650 | 13,876,650 | 579,490 | 635,090 |
| Author FE | Y | Y | Y | | |
| Time Zone FE | Y | Y | Y | | |
| Institution Pair FE | | | | Y | Y |
| Year FE | Y | Y | Y | Y | Y |
| Poisson | | Y | Y | Y | Y |

*Panel B. Interactive Fixed Effects*

| | (1) | (2) | (3) | (4) | (5) |
|---|---|---|---|---|---|
| | Author | | | Institution | |
| | New collab. | Collabo-rators | Publica-tions | Collabo-rators | Publica-tions |
| Overlap × Post | 0.0023 | 0.0334 | 0.0304 | 0.0473 | 0.0341 |
| | (0.0003) | (0.0061) | (0.0056) | (0.0099) | (0.0107) |
| Pretreatment avg. | 0.118 | 1.281 | 0.603 | 3.765 | 2.405 |
| $R^2$ | 0.562 | 0.323 | 0.174 | 0.619 | 0.672 |
| No. of obs. | 8,403,930 | 2,665,838 | 2,665,838 | 579,490 | 635,090 |
| Author × Year FE | Y | Y | Y | | |
| Time Zone × Year FE | Y | Y | Y | Y | Y |
| Institution Pair FE | | | | Y | Y |
| Poisson | | Y | Y | Y | Y |

**Table S1. Alternative FE Specifications:** Table S1 shows that the baseline estimates are robust to both weaker and stronger fixed-effects structures. The baseline specifications absorb time-invariant author heterogeneity together with collaborator-time-zone × year shocks (author design), and institution-pair heterogeneity together with institution-pair × year shocks (institution design). The two-way fixed-effects models in Panel A relax these to additive unit and year effects, assuming a common yearly trend across time zones and across collaborating institutions. The interactive fixed-effects models in Panel B go in the opposite direction: author × year effects absorb every shock to a researcher's total output in a given year, so the estimate compares an author's own high- and low-overlap collaborations within the same year; the institution design analogously combines pair fixed effects with collaborator-time-zone × year effects. Estimates are similar across all three structures, as we would expect if the widening reflects working-hour overlap rather than any particular configuration of unit- or zone-level shocks.

| | (1) | (2) | (3) | (4) | (5) |
|---|---|---|---|---|---|
| | | Author | | Institution | |
| | New collab. | Collabo-rators | Publica-tions | Collabo-rators | Publica-tions |
| *Panel A. Main estimate (2015–2025)* | | | | | |
| Overlap × Post (2021–2025) | 0.0029 | 0.0370 | 0.0279 | 0.0396 | 0.0261 |
| | (0.0009) | (0.0090) | (0.0093) | (0.0078) | (0.0075) |
| *Panel B. Placebo shocks (2010–2019)* | | | | | |
| Overlap × Post (2016–2019) | -0.0017 | -0.0182 | -0.0188 | 0.0016 | -0.0003 |
| | (0.0018) | (0.0149) | (0.0180) | (0.0053) | (0.0099) |
| Overlap × Post (2017–2019) | -0.0019 | -0.0182 | -0.0183 | 0.0001 | 0.0016 |
| | (0.0019) | (0.0148) | (0.0179) | (0.0040) | (0.0097) |
| Overlap × Post (2018–2019) | -0.0019 | -0.0149 | -0.0162 | 0.0009 | 0.0058 |
| | (0.0022) | (0.0152) | (0.0186) | (0.0038) | (0.0089) |
| Pretrend p-value (post 2016–2019) | 0.994 | 0.003 | 0.028 | 0.001 | 0.254 |
| No. of obs. | 5,644,830 | 8,088,120 | 8,088,120 | 418,470 | 469,190 |
| *Panel C. Delayed shocks (2015–2025)* | | | | | |
| Overlap × Post (2022–2025) | 0.0031 | 0.0407 | 0.0304 | 0.0451 | 0.0283 |
| | (0.0010) | (0.0097) | (0.0101) | (0.0084) | (0.0077) |
| Overlap × Post (2023–2025) | 0.0033 | 0.0391 | 0.0296 | 0.0436 | 0.0277 |
| | (0.0011) | (0.0091) | (0.0094) | (0.0073) | (0.0063) |

**Table S2. Placebo Test and Delayed Shocks.** Panel A reproduces the Table 1 estimates. Panel B re-estimates the full design on 2010–2019, with a placebo shock timed at 2015, 2016, or 2017; the estimates are null at every shock year. Panel C re-estimates the main design with the shock delayed to 2022 or 2023, so that work conceived before Covid but published in 2021–22 cannot contribute; the pre-period 2015–2019 is unchanged and 2020 is always omitted. Standard errors, in parentheses, are clustered by time-zone pair.

| | (1) | (2) | (3) |
|---|---|---|---|
| *Panel A. Difference-in-differences, by pre-period team size* | | | |
| | New collab. | Collabo-rators | Publica-tions |
| Large-team authors | 0.0034 | 0.0433 | 0.0392 |
| | (0.0011) | (0.0102) | (0.0084) |
| Small-team authors | 0.0029 | 0.0310 | 0.0224 |
| | (0.0008) | (0.0078) | (0.0060) |
| Pretrend p-value (large) | 0.023 | 0.045 | 0.041 |
| Pretrend p-value (small) | 0.810 | 0.920 | 0.755 |
| No. of obs. (large) | 5,225,310 | 5,225,310 | 5,225,310 |
| No. of obs. (small) | 4,212,720 | 4,212,720 | 4,212,720 |
| *Panel B. Triple difference* | | | |
| | New collab. | Collabo-rators | Publica-tions |
| Overlap × Post | 0.0029 | 0.0344 | 0.0226 |
| | (0.0008) | (0.0086) | (0.0063) |
| Overlap × Post × Large team | 0.0005 | 0.0074 | 0.0166 |
| | (0.0011) | (0.0071) | (0.0063) |
| Gap pretrend p-value | 0.001 | 0.002 | 0.005 |
| No. of obs. | 9,438,030 | 9,438,030 | 9,438,030 |

**Table S3. “Synchronicity advantage” by pre-pandemic team size.** Team size is the author's mean number of authors per paper over 2015–2019; authors who first published after 2019 are not covered. Panel A re-estimates the main author-level difference-in-differences (Overlap × Post) separately within quartile-aligned team-size bins: 1–4, 5–7, 8–10, and more than 10 authors. Panel B estimates a triple difference with the continuous moderator (standardized log pre-period team size), including author and collaborator-time-zone × year fixed effects, team size × year controls, and Overlap × Team size; a team-size-driven widening would appear as a nonzero Overlap × Post × Team size interaction. Standard errors (in parentheses) are clustered by time-zone pair.

| | (1) | (2) | (3) |
|---|---|---|---|
| *Panel A. Difference-in-differences, by author gender* | | | |
| | New collab. | Collabo-rators | Publica-tions |
| Female authors | 0.0025 | 0.0291 | 0.0251 |
| | (0.0007) | (0.0073) | (0.0066) |
| Male authors | 0.0024 | 0.0334 | 0.0250 |
| | (0.0009) | (0.0087) | (0.0080) |
| Pretrend p-value (female) | 0.054 | 0.008 | 0.290 |
| Pretrend p-value (male) | 0.188 | 0.345 | 0.133 |
| No. of obs. (female) | 3,786,120 | 3,786,120 | 3,786,120 |
| No. of obs. (male) | 6,123,020 | 6,123,020 | 6,123,020 |
| *Panel B. Triple difference* | | | |
| | New collab. | Collabo-rators | Publica-tions |
| Overlap × Post | 0.0024 | 0.0322 | 0.0249 |
| | (0.0009) | (0.0086) | (0.0077) |
| Overlap × Post × Female | 0.0002 | -0.0009 | 0.0004 |
| | (0.0007) | (0.0057) | (0.0050) |
| Gap pretrend p-value | 0.282 | 0.613 | 0.384 |
| No. of obs. | 9,909,140 | 9,909,140 | 9,909,140 |

**Table S4. Did the Widening Gap Reflect the Uneven Distribution of Caregiving Responsibilities in Fringe Hours?** Panel A reports the author-design difference-in-differences estimated separately for female and male authors, classified by given name (nomquamgender; female $p \geq 0.8$, male $p \leq 0.2$; 69% of authors classified). Panel B reports a triple difference on the classified sample, with author, collaborator-time-zone × year, and gender × year fixed effects, and Overlap × Female included. If caregiving burdens — which fell disproportionately on women and compete for the fringe hours on which low-overlap collaboration relies — had produced the widening gap, the synchrony advantage would be larger for female authors; the Overlap × Post × Female interaction is instead null for all three outcomes. Standard errors, in parentheses, are clustered by collaborator-time-zone pair.

| | (1) | (2) | (3) |
|---|---|---|---|
| | Author | | |
| | New collab. | Collabo-rators | Publica-tions |
| Overlap × Post | 0.0027 | 0.0336 | 0.0259 |
| | (0.0009) | (0.0086) | (0.0099) |
| Pretreatment avg. | 0.102 | 0.275 | 0.136 |
| $R^2$ | 0.035 | 0.189 | 0.111 |
| No. of obs. | 11,236,820 | 11,236,820 | 11,236,820 |
| Author FE | Y | Y | Y |
| Time Zone × Year FE | Y | Y | Y |
| Poisson | | Y | Y |

**Table S5. Did Authors Relocate Closer to Their Collaborators After COVID?** Results of the author-design regressions restricted to stayers — authors whose home time zone never changes over 2015–2025 (81% of estimation cells). Because a stayer's working-hour overlap with each collaborator time zone is fixed over the study window, the widening gap in this sample cannot reflect authors relocating toward their collaborators. The estimates are essentially unchanged from the full-sample baseline (91–93% of the Table 1 coefficients). Standard errors, in parentheses, are clustered by collaborator-time-zone pair.

| | (1) | (2) | (3) |
|---|---|---|---|
| *Panel A. Authors, by number of collaborator time zones* | | | |
| | New collab. | Collabo- rators | Publica- tions |
| Two or more collaborator time zones | 0.0029 | 0.0343 | 0.0274 |
| | (0.0007) | (0.0079) | (0.0071) |
| One collaborator time zone | 0.0026 | 0.0416 | 0.0274 |
| | (0.0013) | (0.0121) | (0.0141) |
| Pretrend p-value | 0.088 | 0.009 | 0.079 |
| No. of obs. | 5,472,720 | 5,472,720 | 5,472,720 |
| *Panel B. Institution pairs, by researchers' collaborator time zones* | | | |
| | | Collabo- rators | Publica- tions |
| All researchers (Table 2) | — | 0.0396 | 0.0261 |
| | — | (0.0078) | (0.0075) |
| Single-zone researchers only | — | 0.0538 | 0.0542 |
| | — | (0.0116) | (0.0080) |
| Pretrend p-value | — | 0.077 | 0.182 |
| No. of obs. | — | 334,105 | 334,105 |

**Table S6. Single-Time-Zone Author/Institution Event Study.** Panel A estimates the author design separately for authors who collaborated with two or more collaborator time zones over 2015–2025 and for authors who only ever collaborated with one, who cannot shift effort toward higher-overlap ties without finding new collaborators. Panel B re-estimates the institution-pair design with outcomes restricted to researcher dyads both of whose members have a single collaborator time zone; the first row reproduces the Table 1 estimates for comparison. Pretrend p-values and observation counts refer to the single-zone samples. Standard errors, in parentheses, are clustered by time-zone pair in both designs.

| | (1) | (2) | (3) | (4) | (5) |
|---|---|---|---|---|---|
| | Author | | | Institution | |
| | New collab. | Collabo-rators | Publica-tions | Collabo-rators | Publica-tions |
| *Panel A. Overlap × Post* | | | | | |
| Baseline | 0.0031 | 0.0379 | 0.0309 | 0.0474 | 0.0342 |
| | (0.0009) | (0.0097) | (0.0066) | (0.0098) | (0.0107) |
| Controlling for alternatives | 0.0029 | 0.0367 | 0.0286 | 0.0533 | 0.0382 |
| | (0.0009) | (0.0096) | (0.0066) | (0.0106) | (0.0110) |
| No. of obs. | 3,489,390 | 3,489,390 | 3,489,390 | 579,270 | 634,840 |
| *Panel B. Alternatives × Post* | | | | | |
| Zone × year FE | 0.0007 | 0.0036 | 0.0073 | 0.0695 | 0.0475 |
| | (0.0002) | (0.0018) | (0.0015) | (0.0283) | (0.0150) |
| Pretrend p-value | 0.132 | 0.301 | 0.024 | 0.283 | 0.682 |
| No. of obs. | 3,489,390 | 3,489,390 | 3,489,390 | 579,270 | 634,840 |

**Table S7. Interference between a unit's collaborations: Alternatives × Post.** Alternatives is the 2010–2014 portfolio-share-weighted overlap of the unit's other collaborator time zones. Panel A re-estimates the main difference-in-differences controlling for Alternatives × Post; Panel B estimates Alternatives × Post directly, with own Overlap × Post as a regressor and time-zone-pair × year fixed effects. Reallocation of collaboration toward higher-overlap collaborators predicts a negative coefficient.

| | (1) | (2) | (3) | (4) | (5) |
|---|---|---|---|---|---|
| | Author | | | Institution | |
| | New collab. | Collabo-rators | Publica-tions | Collabo-rators | Publica-tions |
| Overlap × Post | 0.0016 | 0.0167 | 0.0188 | 0.0199 | 0.0053 |
| | (0.0007) | (0.0063) | (0.0057) | (0.0058) | (0.0072) |
| Pretreatment avg. | 0.111 | 0.315 | 0.150 | 5.203 | 3.881 |
| $R^2$ | 0.039 | 0.173 | 0.096 | 0.644 | 0.681 |
| No. of obs. | 8,690,950 | 8,690,950 | 8,690,950 | 274,985 | 292,154 |
| Author FE | Y | Y | Y | | |
| Time Zone × Year FE | Y | Y | Y | | |
| Institution Pair FE | | | | Y | Y |
| Institution × Year FE | | | | Y | Y |
| Poisson | | Y | Y | Y | Y |

**Table S8. Did the Widening Gap Reflect Geopolitical Tensions Between the US and China?**

| | (1) | (2) | (3) |
|---|---|---|---|
| | Author | Institution | |
| | Publications | Collaborators | Publications |
| *Panel A. Difference-in-differences, full sample* | | | |
| Overlap × Post | 0.0326 | 0.0309 | 0.0204 |
| | (0.0122) | (0.0053) | (0.0047) |
| Pretreatment avg. | 0.068 | 10.202 | 6.769 |
| Pseudo $R^2$ | 0.253 | 0.590 | 0.619 |
| No. of obs. | 424,930 | 174,737 | 175,620 |
| *Panel B. Triple difference in status asymmetry* | | | |
| Overlap × Post | 0.0330 | 0.0317 | 0.0212 |
| | (0.0123) | (0.0053) | (0.0046) |
| Overlap × Post × Asymmetry | -0.0335 | 0.0189 | 0.0153 |
| | (0.0381) | (0.0153) | (0.0108) |
| Asymmetry × Post | 0.1428 | 0.0812 | 0.0267 |
| | (0.1218) | (0.0469) | (0.0417) |
| Pretreatment avg. | 0.068 | 10.202 | 6.769 |
| Pseudo $R^2$ | 0.254 | 0.590 | 0.619 |
| No. of obs. | 424,930 | 174,737 | 175,620 |
| Pair FE | Y | Y | Y |
| Year FE | Y | | |
| Institution × Year FE | | Y | Y |
| Poisson | Y | Y | Y |

**Table S9. Did the Widening Gap Reflect Status Hierarchy Between Collaborators?**

| | (1) | (2) | (3) | (4) | (5) |
|---|---|---|---|---|---|
| | Author | | | Institution | |
| | New collab. | Collabo-rators | Publica-tions | Collabo-rators | Publica-tions |
| Overlap × Post | 0.0026 | 0.0388 | 0.0248 | 0.0397 | 0.0377 |
| | (0.0014) | (0.0125) | (0.0143) | (0.0075) | (0.0043) |
| Pretreatment avg. | 0.106 | 0.279 | 0.139 | 2.502 | 0.863 |
| $R^2$ | 0.032 | 0.192 | 0.095 | 0.529 | 0.436 |
| No. of obs. | 8,905,840 | 8,905,840 | 8,905,840 | 508,143 | 554,945 |
| Author FE | Y | Y | Y | | |
| Time Zone × Year FE | Y | Y | Y | | |
| Institution Pair FE | | | | Y | Y |
| Institution × Year FE | | | | Y | Y |
| Poisson | | Y | Y | Y | Y |

**Table S10. Excluding collaboration across 3+ time zones**: It's possible in the data that some publication was conducted across more than two time zones, so a further concern is our estimates could be driven by large, multilateral collaborations, which already function with asynchronous arrangement. To address this, we reestimate our main specification excluding all works whose authors (Table S10, columns 1-3) and institutions (columns 4-6) span three and more distinct time zones. The results show that the author-level estimates are basically unchanged, while the institution-pair coefficient rises from 0.0261 to 0.0377, which suggests that our finding is robust.

| | (1) |
|---|---|
| | Publications |
| Overlap × Post | 0.0278 |
| | (0.0112) |
| Pretreatment avg. | 0.342 |
| $R^2$ | 0.245 |
| No. of obs. | 537,310 |
| Author Pair FE | Y |
| Year FE | Y |
| Poisson | Y |

**Table S11. Established Collaborators**: Authors might have formed new collaborations with a new technology in mind even before its rapid expansion due to the pandemic. Zoom, for example, was officially released in 2023. To address the concern of anticipation, we restrict our sample to author pairs with an established coauthorship history between 2010-2014 and reestimate our main DiD specification. This analysis yields a Poisson estimate (+0.0278) nearly identical to the full-sample estimate, which is consistent with the finding at the institution-pair level and confirming that in addition to forming new collaborations, the overlap also impacts the existing collaborations that predate Zoom.

| | (1) | (2) | (3) | (4) | (5) |
|---|---|---|---|---|---|
| | Author | | | Institution | |
| | New collab. | Collabo-rators | Publica-tions | Collabo-rators | Publica-tions |
| High overlap × Post | 0.0099 | 0.1046 | 0.1071 | 0.1037 | 0.0399 |
| | (0.0019) | (0.0355) | (0.0334) | (0.0422) | (0.0397) |
| Pretreatment avg. | 0.119 | 1.311 | 0.628 | 4.230 | 2.598 |
| $R^2$ | 0.569 | 0.323 | 0.173 | 0.651 | 0.696 |
| No. of obs. | 5,117,610 | 1,583,054 | 1,583,054 | 356,680 | 387,537 |
| Author × Year FE | Y | Y | Y | | |
| Time Zone × Year FE | Y | Y | Y | | |
| Institution Pair FE | | | | Y | Y |
| Institution × Year FE | | | | Y | Y |
| Poisson | | Y | Y | Y | Y |

**Table S12. Binary Treatment:** To relax the assumption that the effect of overlap increases linearly, we re-estimate the baseline DiD using “binary treatment” (5–7 hours = high overlap against 0–2 hours = low overlap). Table S6 and Figure S6 report the results. High-overlap collaborations gain on every measure except the number of publications at the institution level: the probability of a new collaboration rises by +0.0099 (SE 0.0019) against a pre-period mean of 0.119, collaborators by 11.0% (+0.105, SE 0.035) and publications by 11.3% (+0.107, SE 0.033). At the institution level, collaborators rise by 10.9% (+0.104, SE 0.042).

| | | (1) | (2) | |
|---|---|---|---|---|
| | | Share of joint publications (%) | Share of institution pairs (%) | |
| 0 h | China–United States | 35.0 | 59.0 | Low overlap |
| | Australia–United States | 11.9 | 3.3 | |
| | Japan–United States | 9.3 | 5.8 | |
| 1 h | Germany–United States | 17.4 | 18.8 | |
| | France–United States | 11.0 | 5.9 | |
| | United Kingdom–Japan | 8.9 | 12.6 | |
| 2 h | China–United Kingdom | 23.3 | 41.7 | |
| | United Kingdom–United States | 20.5 | 4.4 | |
| | Germany–Japan | 5.6 | 5.7 | |
| 3 h | China–Germany | 17.3 | 24.6 | |
| | Germany–United States | 7.3 | 3.8 | |
| | China–Sweden | 5.6 | 6.3 | |
| 4 h | Germany–United States | 19.8 | 20.2 | |
| | France–United States | 11.0 | 6.4 | |
| | United Kingdom–United States | 8.9 | 7.0 | |
| 5 h | United Kingdom–United States | 69.8 | 33.4 | High overlap |
| | Canada–United Kingdom | 13.4 | 5.4 | |
| | Germany–India | 2.5 | 12.5 | |
| 6 h | Brazil–Germany | 13.1 | 4.8 | |
| | India–Japan | 10.6 | 32.8 | |
| | India–South Korea | 9.9 | 26.6 | |
| 7 h | Canada–United States | 45.1 | 17.9 | |
| | Australia–China | 27.2 | 67.5 | |
| | Brazil–United Kingdom | 6.0 | 2.1 | |

**Table S13. Leading Country Pairs at Each Hour of Overlap.**